\documentclass[symmetry,review,accept,pdftex,moreauthors]{Definitions/mdpi} 
\firstpage{1} 
\pubvolume{17}
\issuenum{8}
\articlenumber{1324}
\pubyear{2025}
\copyrightyear{2025}
\externaleditor{Jialun Ping} 
\datereceived{22 July 2025} 
\daterevised{6 August 2025} 
\dateaccepted{8 August 2025} 
\datepublished{14 August 2025} 
\hreflink{https://doi.org/10.3390/\linebreak  sym17081324} 

\usepackage{epsfig}
\usepackage{graphicx}
\usepackage{multirow}
\usepackage{latexsym,amsbsy,amsmath}
\usepackage{stackrel}
\def\llcol#1#2{\tilde{\lambda}_{#1} \cdot \tilde{\lambda}_{#2}}

\Title{Exotic Heavy Hadrons}

\TitleCitation{Exotic Heavy Hadrons}

\Author{Humberto Garcilazo 
 $^{1}$ and Alfredo Valcarce 
 $^{2,}$*\orcidB{}}

\AuthorNames{Humberto~Garcilazo and Alfredo~Valcarce}

\isAPAStyle{%
       \AuthorCitation{Lastname, F., Lastname, F., \& Lastname, F.}
         }{%
        \isChicagoStyle{%
        \AuthorCitation{Lastname, Firstname, Firstname Lastname, and Firstname Lastname.}
        }{
        \AuthorCitation{Garcilazo, H.; Valcarce, A.}
        }
}

\address{%
$^{1}$ \quad Escuela Superior de F\' \i sica y Matem\'aticas, Instituto Polit\'ecnico Nacional, Edificio 9,\linebreak  Mexico City 07738, 
 Mexico; hgarcilazos@ipn.mx\\
$^{2}$ \quad Departamento de F\'\i sica Fundamental, Universidad de Salamanca,  {E-}37008 Salamanca, Spain }

\corres{Correspondence: valcarce@usal.es}

\abstract{We review our recent findings on the structure and properties of exotic heavy hadrons, focusing on two main topics. First, we examine the role of correlations driven by the short-range Coulomb-like color interaction in hidden heavy-flavor pentaquarks. We show how this framework consistently accounts for the observed pattern of $P_c$ and $P_{cs}$ states in the hidden-charm sector and enables predictions for the hidden-bottom sector, where experimental data are still lacking. The second topic explores the possibility of forming stable multihadron molecules from deeply bound two-hadron exotic states. In this context, a bound state of three $B$ mesons, denoted as $T_{bbb}$, with quantum numbers $(I)J^P = (1/2)2^-$, is presented. We find that the binding energy generally decreases as the number of hadrons increases, primarily due to effects of the Pauli principle and the appearance of new decay thresholds. Nonetheless, resonances may still arise in specific cases, depending on the internal thresholds of the system. Finally, we discuss how the decay width of an exotic multihadron resonance can offer valuable insights into its internal structure and underlying~dynamics.}


\begin{document}

\section{Introduction}
\label{secI}
Hadronic spectroscopy has become an essential tool for improving our 
understanding of Quantum Chromodynamics (QCD) in the low-energy regime. 
Broadly speaking, at the beginning of the 21st century, two distinct 
approaches coexisted for describing the phenomenology of the strong 
interaction in this domain.

On one hand, there was the traditional approach to hadronic 
spectroscopy, which focused on studying the spectra of mesons 
and baryons in the spirit of Gell-Mann's model~\cite{Gel64}, 
where hadrons are treated as systems of constituent (anti)quarks. 
However, it soon became clear that certain states could not be 
easily accommodated within this simple picture. 
Examples include mesons with explicitly exotic quantum 
numbers that cannot be described as a simple quark--antiquark pair, 
such as the $\pi_1(1400)$~\cite{Bar95}, or the long-standing
puzzle of the scalar 
mesons~\cite{Jaf77,Jae77}. Anomalies have also emerged in the baryon
sector, with early indications of more complex dynamics. One 
prominent example has been the $\Lambda(1405)$, 
whose mass lies significantly below
the predictions of many quark models~\cite{Vac05}. This discrepancy
led to the theoretical prediction of a double-pole structure~\cite{Mag05},
although some models interpret it as a conventional three-quark 
state~\cite{Cap86}. Predictions of dibaryon
candidates have also been made, though in some cases there is still no
consensus regarding their
internal structure, whether they are bound states of two baryons
or genuine hexaquark configurations~\cite{Cle17}. Finally, one cannot
overlook the controversial case of the $\Theta^+ (1540)$~\cite{Tri04}. 
From a theoretical perspective, even relatively simple models that
captured the essential features of QCD had already predicted the
possible existence of more 
complex configurations, know as multiquarks~\cite{Ade82}.

On the other hand, an alternative view was based on the use of
asymptotic hadronic states, effective field theories (EFTs)~\cite{Pic95}. 
EFTs provide a suitable theoretical framework for describing low-energy 
physics, where ``low'' is defined relative to a
characteristic energy scale $\Lambda$. 
These theories retain only the relevant degrees of 
freedom and 
states with mass $m \ll \nolinebreak \Lambda$, 
while heavier states with $m \gg \Lambda$ 
are systematically integrated out. 
This leads to a non-renormalizable theory 
organized as a power expansion in energy ($\Lambda$). EFTs have been 
successfully employed to describe hadronic interactions and to 
explore the emergence of 
higher-mass resonances beyond the effective theory's cutoff~\cite{Oll00}.

Over the past two decades, the number of newly observed states in the 
heavy-hadron spectrum has increased significantly. 
This has lead to the growing consensus 
that the heavy-hadron spectrum includes 
contributions from configurations beyond the simplest quark--antiquark 
(meson) or three-quark (baryon) structures proposed by 
Gell-Mann~\cite{Gel64}. This is particularly evident in the recent 
discovery of doubly heavy tetraquarks with manifestly exotic quantum 
numbers~\cite{Aai22,Aaj22}. Nonetheless, many of the newly reported 
states possess ordinary quantum numbers, suggesting that they may
arise from more intricate internal 
structures allowed by QCD~\cite{Jaf05}.

These experimental advances have sparked substantial theoretical 
efforts aimed at understanding the spectroscopy and 
internal structure of these novel 
states~\cite{Che16,Bri16,Ric16,Hos16,Che17,Leb17,Ali17,Esp17,Guo18,Ols18,Kar18,Bra20,Yan20,Hua23}. 
A wide range of models has been proposed, including hadronic molecules, 
diquarks, hadroquarkonium, hybrids, threshold effects, \dots, 
each providing valuable insights yet none offering 
a complete description. No single framework has succeeded 
in fully accounting for all 
the observed states. A comprehensive understanding may require 
the interplay of multiple mechanisms, with different combinations
being relevant for different cases.

A central open question in this context is whether states with genuinely 
complex quark substructures, so-called multiquarks, can be unambiguously 
identified. For hadrons with manifestly exotic quantum 
numbers, recent experimental discoveries~\cite{Aai22,Aaj22} provide a 
clear affirmative answer. However, for states 
with non-exotic quantum numbers, 
the question remains unresolved~\cite{Jaf77,Jae77}. A striking 
feature of non-conventional hadrons with standard quantum numbers is the 
rarity of bound states, which appear only in very specific
configurations. This observation is supported by both lattice QCD 
simulations~\cite{Hug18,Hud20,Col24} and constituent quark 
models~\cite{Sil93,Ric18}. 

In Section~\ref{secII}, we will address the observed difficulty 
in forming stable multiquark 
states that do not undergo prompt decay, which may signal the presence 
of nontrivial internal correlations, dynamical constraints that 
restrict the 
allowed quantum numbers of their constituent subsystems.
Multiquark systems (tetraquarks, pentaquarks, etc.) 
necessarily contain 
internal color-singlet components. Nevertheless,
in contrast to atomic or nuclear physics, their 
dominant Fock-space components may correspond to 
color-connected configurations that are not asymptotically
separable into isolated hadrons. This 
opens the possibility for theoretical frameworks 
in which multiquark dynamics is 
governed by QCD-induced correlations~\cite{Jaf05}. 
Within this context, we 
recently investigated hidden-flavor pentaquarks by examining the 
correlations arising from the Coulomb-like behavior of the short-range 
color interaction~\cite{Gar22,Gac22}. Our analysis yields 
a spectral pattern that aligns 
with current experimental observations.

The ongoing theoretical efforts in hadron spectroscopy have 
also underscored another important aspect of low-energy QCD
phenomenology: the prediction of stable 
two-baryon states containing multiple heavy 
quarks, as suggested both by lattice QCD 
simulations~\cite{Gon18,Jua19,Lyu21,Mat23} and by 
various phenomenological models~\cite{Gac18,Mam19,Wul23,Gar24,Gar25}. 
Nonetheless, several of these predictions 
have been challenged by rigorous few-body calculations, which 
argue against the formation of such bound states~\cite{Ric20}. 

Section~\ref{secIII} is motivated by theoretical predictions of 
deeply bound states involving heavy flavors,
specifically the doubly bottom tetraquark 
$T_{bb}$~\cite{Hud20,Col24} and
the $\Omega_{bbb} \Omega_{bbb}$ dibaryon~\cite{Mat23}.
Thus, we review the viability of forming multihadron molecular states
composed of such heavy-flavor 
constituents~\cite{,Wul23,Gar25}. 
Our analysis indicates that several dynamical 
effects contribute to a progressive reduction in 
the binding energy as the number of hadrons increases, 
thereby disfavoring the existence 
of stable multihadron configurations with heavy quarks. 

In Section~\ref{secIV}, we will address the properties of
resonances that often emerge near the various 
physical thresholds intrinsic to the multiquark system 
given the inherent difficulty in
forming bound multiquark states.
One particularly intriguing phenomenon in this context 
is the appearance of narrow resonances 
despite the presence of ample phase space for decay. 
This seemingly paradoxical behavior can be understood 
as a consequence of the orthogonality between 
the color wave functions of the physical thresholds,
which effectively provides a stabilizing mechanism~\cite{Gar18,Gai22}.

Finally, Section~\ref{secV} offers a concise summary of our 
main findings.

\section{Hidden-Flavor Pentaquarks}
\label{secII}

Table~\ref{Tab_1} summarizes the currently known hidden-charm 
pentaquarks, including both nonstrange and strange
candidates~\cite{Aai15,Aai19,Che22,Aak21}. 
The first four entries, collectively referred to as $P_c$ states, are 
consistent with a minimal quark configuration of $c\bar{c}qqq$, 
where $q = u$ or $d$~\cite{Aai15,Aai19}. The subsequent entries
correspond to candidates for strange hidden-charm pentaquarks,
characterized by a minimal quark content of $c\bar{c}uds$.

\begin{table}[H]
\caption{{Summary} 
of established $P_c^+$ and $P_{cs}$ pentaquark candidates~\cite{Aai15,Aai19,Che22,Aak21}.}\label{Tab_1} 
\begin{tabularx}{\textwidth}{CCC}
\toprule
 \textbf{State}                                   & \textbf{\emph{M} (MeV) }                       & \boldmath{$\Gamma$} \textbf{(MeV)   }               \\ \midrule
 $P_c(4312)^+$                           & $4311.9 \pm 0.7 ^{+6.8}_{-0.6}$  & $9.8 \pm 2.7 ^{+3.7}_{-4.5}$     \\ 
 $P_c(4380)^+$                           & $4380 \pm 8 \pm 29$              & $205 \pm 18 \pm 86$              \\
 $P_c(4440)^+$                           & $4440.3 \pm 1.3 ^{+4.1}_{-4.7}$  & $20.6 \pm 4.9 ^{+8.7}_{-10.1}$   \\ 
 $P_c(4457)^+$                           & $4457.3 \pm 0.6 ^{+4.1}_{-1.7}$  & $6.4 \pm 2.0 ^{+5.7}_{-1.9}$     \\ \midrule
$P_{\Psi s}^\Lambda(4338)$              & $4338.2 \pm 0.7 \pm 0.4$         & $7.0 \pm 1.2 \pm 1.3$            \\ 
 $P_{cs}(4459)$                          & $4458.8 \pm 2.9 ^{+4.7}_{-1.1}$  & $17.3 \pm 6.5 ^{+8.0}_{-5.7}$    \\
{$P_{cs}(4459)$}~\textsuperscript{1} & $4454.9 \pm 2.7$                 & $7.5 \pm 9.7$                    \\ 
                                       &  $4467.8 \pm 3.7$                & $5.2 \pm 5.3$                    \\ 
\bottomrule
\end{tabularx}
\noindent{\footnotesize{\textsuperscript{1}{Alternative} 
assignment of Ref.~\cite{Aak21}.}}

\end{table}
\vspace{-6pt}

Recent reviews have extensively covered both experimental results and 
theoretical interpretations of the nonstrange sector~\cite{Che16,Bri16,Ric16,Hos16,Che17,Leb17,Ali17,Esp17,Guo18,Ols18,Kar18,Bra20}. 
More recently, the LHCb Collaboration reported the observation of 
a new strange pentaquark, $P_{\Psi s}^\Lambda(4338)$, identified as a 
resonance in the $J/\psi \, \Lambda$ invariant mass spectrum 
from the decay $B^- \to J/\psi \Lambda \bar{p}$~\cite{Che22}. This state 
has a mass of $4338.2 \pm 0.7 \pm 0.4$~MeV.
Additionally, LHCb presented evidence for another structure in the 
$J/\psi \, \Lambda$ invariant mass distribution from an 
amplitude analysis of $\Xi_b^- \to J/\psi \, \Lambda K^-$ 
decays~\cite{Aak21}. This feature 
appears at $4458.8 \pm 2.9^{+4.7}_{-1.1}$~MeV, although the data are also 
consistent with two nearby resonances at masses of 
$4454.9 \pm 2.7$~MeV and \mbox{$4467.8 \pm 3.7$~MeV}. These alternative 
interpretations are reflected in the final entry of Table~\ref{Tab_1}. If 
confirmed, the existence of two strange pentaquarks would parallel the 
pattern observed in the nonstrange sector, exemplified by the 
$P_c(4440)^+$ and $P_c(4457)^+$ states.

\textls[-20]{It is worth noting that recent studies by the Belle Collaboration
rule out the production of $P_c^+$ pentaquarks in $\Upsilon(1,2S)$
decays~\cite{Don24} but provide evidence for the existence of $P_{cs}$ 
pentaquarks~\cite{Ada25}. In particular, for the $P_{cs}(4459)$ state,
Belle reports a mass of \mbox{$4471.7 \pm 4.8 \pm 0.6$ MeV/c$^2$}.}

\subsection{Coulomb-like Color Correlations}

\label{subsecII.I}

As noted earlier, a central theoretical challenge in the study of
multiquarks is identifying multiquark hadrons that do not promptly decay
through their fall-apart channel into lighter hadrons. This challenge
has motivated proposals emphasizing internal correlations, 
dynamical mechanisms that constrain the allowed quantum numbers of
subsystems, thereby enhancing stability. 
Notable among these are diquark models~\cite{Ans93,Fre82}, 
which restrict the color degree of freedom of quark pairs 
to specific configurations.

\textls[-15]{Various diquark models exist, but all impose constraints on the color 
states of quark--quark or quark--antiquark pairs. For example, some 
models require diquarks to be in a {color} 
 $\mathbf{\bar{3}}$ 
state~\cite{Mai15,Gir19,Ali19,Shi21}, while others focus on 
quark--antiquark pairs restricted to a 
color $\mathbf{8}$~\cite{Wul17}. If a 
multiquark contains color structures not found 
in asymptotic hadron states, 
this could act as a stabilizing mechanism, enhancing the possibility 
of bound-state formation. Compared to uncorrelated models, 
these approaches generally predict a richer spectrum of exotic hadrons.}

\begin{figure}[H]
\includegraphics[width=0.45\columnwidth]{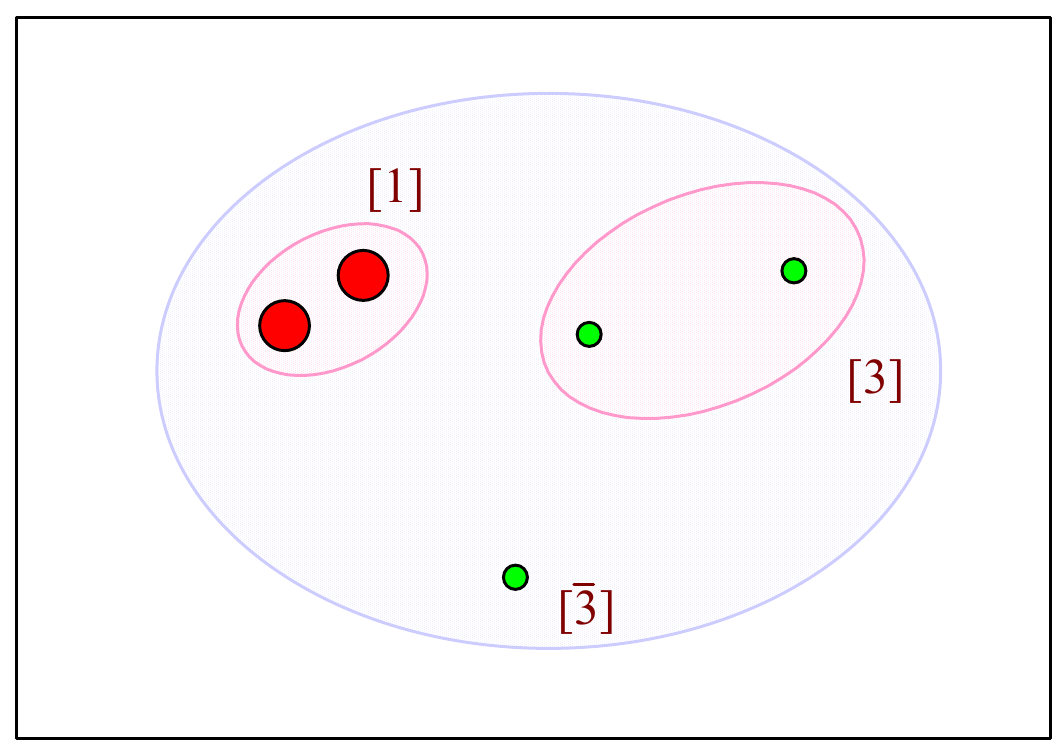}
\caption{{Color} 
structure of a hidden-flavor pentaquark driven by 
Coulomb-like color correlations between the heavy quarks. Large red 
circles denote the heavy quark--antiquark ($Q\bar{Q}$) pair,
while small green circles represent the light 
quarks. Bracketed numbers indicate the corresponding
color quantum numbers.}
\label{Fig_1}
\end{figure}

We have employed dynamical correlations induced by 
the Coulomb-like short-range color interaction between heavy quarks
to study hidden-flavor pentaquarks of the form 
$Q\bar{Q}qqq'$, where $Q = b$ or $c$, $q = u$ or $d$, and 
$q' = u, d, s$~\cite{Gar22,Gac22}. 
While traditional diquark models often emphasize 
attractive $qq$ interactions in a 
color-$\mathbf{\bar{3}}$ state, we argue that the $Q\bar{Q}$ color 
singlet is even more favorable due to binding energy scaling with the 
heavy quark mass $M_Q$. 
In natural units, the binding energy of a $Qq$ diquark scales
with the light quark mass $m_q$, whereas that of the
color-singlet $Q\bar{Q}$ pair scales as $2M_Q$.
This favors an internal structure dominated by a color-singlet $Q\bar{Q}$ 
pair rather than a color-octet~\cite{Jaf05}, as illustrated in 
Figure~\ref{Fig_1}. Taking into account the isospin-zero nature 
of the heavy quarks and the antisymmetry of the 
color-$\mathbf{\bar{3}}$ $qq$ wave function 
(which enforces matching spin and isospin), we construct the wave
function for the lowest-lying states, that is, those with a fully 
symmetric radial wave function, of any $(I,J)$ pentaquark:
\begin{equation}
\Psi^{(I,J)}_{\rm Pentaquark} \,\, = \,\, \{{\bf 3}_c, i_1, s_1=1/2 \}_{q} \, \otimes \, 
\{{\bf 1}_c, i_2=0, s_2 \}_{(Q \bar Q)} \, \otimes \, 
\{{\bf \bar 3}_c, i_3=s_3, s_3\}_{(qq)} \, ,
\label{Equ_1}
\end{equation}
\textls[-20]{where $i_n$ and $s_n$ denote the the isospin and spin of the different 
components. $i_1 = 1/2$ for $Q\bar{Q}qqq$ 
and $i_1 = 0$ for $Q\bar{Q}qqs$. 
The Coulomb-like correlations between the heavy quarks help to avoid 
repulsive interactions associated with color-octet $Q\bar{Q}$ 
configurations. Notably, these constraints also arise naturally in 
constituent quark models for doubly heavy tetraquarks~\cite{Her20,Men21}.}
\vspace{-9pt}

QCD favors certain quark configurations over others~\cite{Jaf05}. 
As discussed above, the strongest quark--antiquark correlation 
occurs in the color-, flavor-, and spin-singlet channel
$\{{\mathbf{1}_c, \mathbf{1}_f, 0_s}\}$. 
The next most favorable configuration is the 
color-antitriplet, flavor-antisymmetric, spin-singlet channel 
$\{{\mathbf{\bar{3}}_c, \mathbf{\bar{3}}_f, 0_s}\}$, which 
underlies the structure of diquarks. 
Table~\ref{Tab_2} lists hidden-flavor pentaquark 
states containing at least one such favorable QCD channel, i.e., a 
spin-zero diquark.
\vspace{-3pt}

\begin{table}[H]
\caption{{Quantum} 
numbers of $Q\bar{Q}qqq$ hidden-flavor pentaquarks  
containing spin-zero diquarks.
Numbers in parentheses denote the isospin 
of $Q\bar{Q}qqs$ pentaquarks. See Equation~(\ref{Equ_1}) for notation.} \label{Tab_2}
\begin{tabularx}{\textwidth}{CCCCCC}
\toprule
        \boldmath{$I$}           & \boldmath{$J$}     & \boldmath{$s_1$}   & \boldmath{$s_2$ }   & \boldmath{$s_3$} & \textbf{Vector} \\ \midrule
{$1/2 \, (0)$}   & $1/2$   & {$1/2$}   & $0$      & $0$    &  $v_1$ \\
                         & $1/2$   & & $1$      & $0$    &  $v_2$ \\
                         & $1/2$   &    & $0$      & $1$    &  $v_3$ \\
                         & $3/2$   &    & $1$      & $0$    &  $w_1$ \\ 	\midrule
$3/2 \, (1)$                    & $3/2$   &$1/2$   & $0$      & $1$    &  $w_3$ \\
\bottomrule
\end{tabularx}
\end{table}
\vspace{-6pt}

\textls[-15]{To analyze the effects of Coulomb-like correlations on the
structure of hidden-flavor pentaquarks, we 
employ the generic constituent quark model known as the 
AL1 potential~\cite{Sem94}}. This 
model has been widely used in studies of 
multiquark systems~\cite{Sil93,Ric18,Jan04,Her20,Ric17,Hiy18,Meg19}. It 
features a central potential combining a Coulomb term with a
linear confinement term, complemented by 
a smeared chromomagnetic spin--spin interaction:
\begin{align}
\label{Equ_2}
V(r)  & =  -\frac{3}{16}\, \llcol{i}{j}
\left[\lambda\, r - \frac{\kappa}{r}-\Lambda + \frac{V_{SS}(r)}{m_i \, m_j}  \, \vec \sigma_i \cdot \vec\sigma_j\right] \, ,\\ \nonumber \\
V_{SS}  &= \frac{2 \, \pi\, \kappa^\prime}{3\,\pi^{3/2}\, r_0^3} \,\exp\left(- \frac{r^2}{r_0^2}\right) ~,\quad
 r_0 =  A \left(\frac{2 m_i m_j}{m_i+m_j}\right)^{-B} \, . \nonumber
\end{align}

The model parameters are set as follows: 
$\lambda = 0.1653~\mathrm{GeV}^2$, $\Lambda = 0.8321~\mathrm{GeV}$, 
\mbox{$\kappa = 0.5069$}, $\kappa' = 1.8609$, $A = 1.6553~\mathrm{GeV}^{B-1}$, 
and $B = 0.2204$. The constituent quark masses are as follows: 
$m_u = m_d = 0.315~\mathrm{GeV}$, $m_s = 0.577~\mathrm{GeV}$, 
$m_c = 1.836~\mathrm{GeV}$, and $m_b = 5.227~\mathrm{GeV}$. The color 
factor $\llcol{i}{j}$ is adapted to differentiate between 
quark--quark and quark--antiquark interactions.

\textls[-15]{Notably, the smearing of the spin--spin interaction 
depends on the reduced mass 
of the interacting quark pair, thereby maintaining the flavor dependence 
of the interaction. The AL1 potential parameters were finely tuned 
through a global fit to 36 mesons and 53 baryons, achieving excellent 
agreement with experimental data, as detailed in Table 2 
of Ref.~\cite{Sem94}. 
Although the reported $\chi^2$ value is slightly
higher than that of some other models, this is largely due 
to the inclusion of high-angular-momentum resonances. Nevertheless, 
the AL1 model remains highly effective for analyzing low-energy hadron 
spectra~\cite{Sil96}. The color--spin algebra pertinent to five-quark 
systems has been systematically developed in Refs.~\mbox{\cite{Ale11,Ric17}}.}

In multiquark dynamics, binding can arise either from medium- to 
long-range attraction, often modeled through Goldstone boson exchange 
between color-singlet \linebreak clusters~\mbox{\cite{Vij04,Hua16,Yan17}}, or from 
short-range dynamics driven by one-gluon exchange. While the latter 
typically generates strong repulsion in nucleon--nucleon ($NN$) $S$-waves, 
it does not universally apply to all hadronic channels. For example, 
certain configurations in the $\Delta\Delta$ and $N\Delta$ sectors 
display short-range attraction, as evidenced by their low-energy phase 
shifts~\cite{Oka80}. This attraction has underpinned
predictions of resonances in 
these \linebreak systems~\cite{Gol89,Pan01,Val01,Val05}, 
some of which may have experimental support~\cite{Cle17}. 
For a detail discussion on
the interplay between Goldstone boson and one-gluon exchange mechanisms within hybrid 
constituent models, particularly regarding 
tetraquark stability, see Refs.~\cite{Vij04,Saz22}.

\subsection{The Three-Body Problem}
\label{subsecII.III}

Multiquark systems may exhibit internal clustering, which can
significantly 
simplify their computational treatment. This is especially evident in 
tetraquark studies, where a diquark--antidiquark approximation
reduces the original four-body problem to an effective 
two-body one~\cite{Kar17,Eic17}. For hidden-flavor pentaquarks, 
the factorization of color degrees of freedom (as 
discussed in Section~\ref{subsecII.I}), combined with the flavor 
independence of the quark--quark interaction, 
reduces the five-body problem to an 
effective three-body system, as illustrated in Figure~\ref{Fig_1}. This 
effective system can be solved exactly using Faddeev
equations~\cite{Fad61,Fad65}, thereby 
avoiding the convergence difficulties often encountered
with variational methods, particularly near thresholds. 
Our numerical approach follows the 
formalism outlined in Ref.~\cite{Gar03}.

The structure of the Faddeev equations is governed by conservation 
laws. In particular, three-body states in which 
a particle has a fixed 
spin can only couple to other states with the same spin due to 
the orthogonality of spinors. An analogous rule 
applies to isospin. Consequently, the integral equations decouple 
into separate sets, each 
characterized by conserved individual spin and isospin values. The sets 
relevant to total spin $J=1/2$ and $J=3/2$ are presented in 
Table~\ref{Tab_4}. Here,
$S_i$ denotes the spin of the pair $jk$. 
Because the individual isospin is determined 
by the spin configuration, it is 
not explicitly listed in the tables. The matrix element 
$F = \langle \vec \sigma_i \cdot \vec \sigma_j \rangle$ represents the 
coupling strength between two-body amplitudes mediated by the spin--spin 
interaction.

\begin{table}[H]
\small
\caption{{Channels} 
contributing to $J=1/2$ and $J=3/2$
states, denoted by $v_i$ and $w_i$, respectively, in~Table~\ref{Tab_2}.}
\begin{tabularx}{\textwidth}{CCCCCCC}
\toprule
\multicolumn{7}{c}{\boldmath{$J=1/2$}} \\ 
\midrule
                      & $s_1$ & $s_2$ & $S_3$    & $s_3$& $ I $     & $ F $           \\ \midrule
$v_1$                 & $1/2$ & $ 0 $ & $ 1/2  $ & $ 0 $& $1/2$     & $9/8$           \\
$v_2$                 & $1/2$ & $ 1 $ & $ 1/2  $ & $ 0 $& $1/2$     & $3/8$           \\
$v_3$                 & $1/2$ & $ 0 $ & $ 1/2  $ & $ 1 $& $1/2$     & $9/8$           \\
\cmidrule{1-5}
                      & $s_2$ & $s_3$ & $S_1$    & $s_1$&           &           \\ \cmidrule{1-5}
$v_1$                 & $ 0 $ & $ 0 $ & $  0   $ & $1/2$& $1/2$     & $ 0 $      \\
$v_2$                 & $ 1 $ & $ 0 $ & $  1   $ & $1/2$& $1/2$     & $ 0 $      \\
$v_3$                 & $ 0 $ & $ 1 $ & $  1   $ & $1/2$& $1/2$     & $ 0 $      \\
\cmidrule{1-5}
                      & $s_3$ & $s_1$ & $S_2$    & $s_2$&           &                \\ \cmidrule{1-5}
$v_1$                 & $ 0 $ & $1/2$ & $ 1/2  $ & $ 0 $& $1/2$     & $9/8$           \\
$v_2$                 & $ 0 $ & $1/2$ & $ 1/2  $ & $ 1 $& $1/2$     & $9/8$           \\
$v_3$                 & $ 1 $ & $1/2$ & $ 1/2  $ & $ 0 $& $1/2$     & $3/8$           \\
 \midrule
 \multicolumn{7}{c}{\boldmath{$J=3/2$}}   \\ 
\midrule
                      & $s_1$ & $s_2$ & $S_3$    & $s_3$& $ I $     & $ F $           \\ \midrule
$w_1$                 & $1/2$ & $ 1 $ & $ 3/2  $ & $ 0 $& $1/2$     & $3/2\sqrt{2}$    \\
$w_3$                 & $1/2$ & $ 0 $ & $ 1/2  $ & $ 1 $& $3/2$     & $9/8$           \\										\cmidrule{1-5}	
                      & $s_2$ & $s_3$ & $S_1$    & $s_1$&           &            \\ \cmidrule{1-5}
$w_1$                 & $ 1 $ & $ 0 $ & $  1   $ & $1/2$& $1/2$     & $ 0 $      \\
$w_3$                 & $ 0 $ & $ 1 $ & $  1   $ & $1/2$& $3/2$     & $ 0 $      \\		
		\cmidrule{1-5}						
                      & $s_3$ & $s_1$ & $S_2$    & $s_2$&           &                \\ \cmidrule{1-5}
$w_1$                 & $ 0 $ & $1/2$ & $ 1/2  $ & $ 1 $& $1/2$     & $9/8$           \\
$w_3$                 & $ 1 $ & $1/2$ & $ 3/2  $ & $ 0 $& $3/2$     & $3/2\sqrt{2}$   \\											
\bottomrule
 
\end{tabularx}
\label{Tab_4} 
\end{table}
\vspace{-6pt}

We briefly review the solution of the Faddeev equations in the case 
where the three~constituents are in $S$-wave configuration. The
generalization to states with orbital angular momentum $L\neq 0$ 
results in a coupled-channel problem, as 
detailed in Ref.~\cite{Gar22}. Accordingly,
the Faddeev equations for a 
bound state with total isospin $I$ and total spin $J$ 
can be expressed as
\begin{eqnarray}
T_{i;IJ}^{I_iS_i}(p_iq_i) = &&\sum_{j\ne i}\sum_{I_jS_j}
\frac{1}{2}\int_0^\infty q_j^2dq_j
\int_{-1}^1d{\rm cos}\theta\, 
t_{i;I_iS_i}(p_i,p_i^\prime;E-q_i^2/2\nu_i) 
\nonumber \\ &&
\times \,
h_{ij;IJ}^{I_iS_i;I_jS_j}
\frac{1}{E-p_j^2/2\eta_j-q_j^2/2\nu_j}\;
T_{j;IJ}^{I_jS_j}(p_jq_j) \, , 
\label{eq1} 
\end{eqnarray}
where $t_{i;I_iS_i}$ are the two-body scattering amplitudes, and $\eta_i$, $\nu_i$ are the reduced masses:
\begin{eqnarray}
\eta_i &=& \frac{m_jm_k}{m_j+m_k} \, , \nonumber\\
\nu_i &=& \frac{m_i(m_j+m_k)}{m_i+m_j+m_k} \, .
\label{eq3}
\end{eqnarray}
%

$\vec p_i^{\; \prime}$
is the momentum of the pair $jk$ (with $ijk$ an even permutation of
$123$) and $\vec p_j$ is the momentum of the pair
$ki$ which are given by
\begin{eqnarray}
\vec p_i^{\; \prime} &=& -\vec q_j-\alpha_{ij}\vec q_i \, , \nonumber\\
\vec p_j &=& \vec q_i+\alpha_{ji}\vec q_j \, ,
\label{eq3p}
\end{eqnarray}
where
\begin{eqnarray}
\alpha_{ij} &=& \frac{\eta_i}{m_k} 
\, , \nonumber\\
\alpha_{ji} &=& \frac{\eta_j}{m_k} 
\, ,
\label{eq3pp}
\end{eqnarray}
so that
\begin{eqnarray}
p_i^\prime &=& \sqrt{q_j^2+\alpha_{ij}^2q_i^2+2\alpha_{ij}
q_iq_j{\rm cos}\theta} \, , \nonumber \\
p_j &=& \sqrt{q_i^2+\alpha_{ji}^2q_j^2+2\alpha_{ji}
q_iq_j{\rm cos}\theta} \, .
\label{eq5}
\end{eqnarray}
%

$h_{ij;IJ}^{I_iS_i;I_jS_j}$ are the spin--isospin coefficients,
\begin{eqnarray}
h_{ij;IJ}^{I_iS_i;I_jS_j}= &&
(-)^{I_j+i_j-I}\sqrt{(2I_i+1)(2I_j+1)}
W(i_ji_kIi_i;I_iI_j)
\nonumber \\ && \times
(-)^{S_j+s_j-J}\sqrt{(2S_i+1)(2S_j+1)}
W(s_js_kJs_i;S_iS_j) \, , 
\label{eq6}
\end{eqnarray}
where $W$ is a Racah coefficient and $i_i$, $I_i$, and $I$ 
($s_i$, $S_i$, and $J$) are the isospins (spins) of particle $i$
of the pair $jk$ and of the three-body system.

To facilitate numerical treatment, we map the momentum variable 
$p_i \in [0, \infty)$ to $x_i \in [-1, 1]$ via

\begin{equation}
x_i=\frac{p_i-b}{p_i+b} \, ,
\label{eq7}
\end{equation}
where $b$ is an arbitrary scaling parameter. This transforms 
the Faddeev equations into
\begin{eqnarray}
T_{i;IJ}^{I_iS_i}(x_iq_i) = &&\sum_{j\ne i}\sum_{I_jS_j}
\frac{1}{2}\int_0^\infty q_j^2dq_j
 \int_{-1}^1d{\rm cos}\theta\; 
t_{i;I_iS_i}(x_i,x_i^\prime;E-q_i^2/2\nu_i) 
\nonumber \\ &&
\times \,
h_{ij;IJ}^{I_iS_i;I_jS_j}
\frac{1}{E-p_j^2/2\eta_j-q_j^2/2\nu_j}\;
T_{j;IJ}^{I_jS_j}(x_jq_j) \, . 
\label{eq8} 
\end{eqnarray}

The kernel can be further simplified by expanding the 
two-body amplitude in Legendre polynomials:
\begin{equation}
t_{i;I_iS_i}(x_i,x_i^\prime;e)=\sum_{nr}P_n(x_i)\tau_{i;I_iS_i}^{nr}(e)P_r(x'_i) \, ,
\label{eq9}
\end{equation}
where the expansion coefficients are given by
\begin{equation}
\tau_{i;I_iS_i}^{nr}(e)= \frac{2n+1}{2}\frac{2r+1}{2}\int_{-1}^1dx_i
\int_{-1}^1 dx_i^\prime\; P_n(x_i) 
t_{i;I_iS_i}(x_i,x_i^\prime;e)P_r(x_i^\prime) \, .
\label{eq10} 
\end{equation}

Substituting this expansion yields
\begin{equation}
T_{i;IJ}^{I_iS_i}(x_iq_i) = \sum_n P_n(x_i) T_{i;IJ}^{nI_iS_i}(q_i) \, ,
\label{eq11}
\end{equation}
where $T_{i;IJ}^{nI_iS_i}(q_i)$ satisfies the one-dimensional integral equation
\begin{equation}
T_{i;IJ}^{nI_iS_i}(q_i) = \sum_{j\ne i}\sum_{mI_jS_j}
\int_0^\infty dq_j K_{ij;IJ}^{nI_iS_i;mI_jS_j}(q_i,q_j;E)\;
T_{j;IJ}^{mI_jS_j}(q_j) \, , 
\label{eq12}
\end{equation}
with
\begin{eqnarray}
K_{ij;IJ}^{nI_iS_i;mI_jS_j}(q_i,q_j;E)= &&
\sum_r\tau_{i;I_iS_i}^{nr}(E-q_i^2/2\nu_i)
\frac{q_j^2}{2}
\nonumber \\ &&
\times\int_{-1}^1 d{\rm cos}\theta\;
h_{ij;IJ}^{I_iS_i;I_jS_j}
\frac{P_r(x_i^\prime)P_m(x_j)} 
{E-p_j^2/2\eta_j-q_j^2/2\nu_j} \, .
\label{eq13} 
\end{eqnarray}

The three amplitudes $T_{1;IJ}^{rI_1S_1}(q_1)$, $T_{2;IJ}^{mI_2S_2}(q_2)$,
and $T_{3;IJ}^{nI_3S_3}(q_3)$ in Equation~(\ref{eq12}) are coupled together.

\subsection{Results and Discussion}
\label{subsecII.IV}
We solved the three-body problem for various $(I, J)$ configurations 
of the $Q\bar{Q}qqq'$ pentaquark systems discussed in 
Section~\ref{subsecII.I}. The resulting binding energies 
for the different hidden-flavor pentaquark 
configurations are summarized in Table~\ref{Tab_5}.

\begin{table}[H]
\caption{{Binding} 
 energies (in MeV) of the different $Q\bar{Q}qqq'$ pentaquark states.}
\begin{tabularx}{\textwidth}{CCCCCCCC} 
\toprule
\boldmath{$Q$} & \boldmath{$q$}   & \boldmath{$q'$ } & \boldmath{$v_1$ }& \boldmath{$v_2$} & \boldmath{$v_3$} & \boldmath{$w_1$ }& \boldmath{$w_3$} \\ \midrule
$c$ & $u,d$ & $u,d$ & 7     & 17    & 24    & 12    & 12    \\
$c$ & $u,d$ & $s$   & 133   & 138   & 143   & 134   & 134   \\
$b$ & $u,d$ & $u,d$ & 39    & 41    & 52    & 40    & 40   \\
$b$ & $u,d$ & $s$   & 165   & 167   & 175   & 166   & 166 \\
\bottomrule
\end{tabularx}
\label{Tab_5}
\end{table}
\vspace{-3pt}

We first observe an apparent degeneracy between the 
$w_1$ and $w_3$ vectors.
These correspond to $I = 1/2$ and $I = 3/2$ $Q\bar{Q}qqq$ pentaquarks
with $J = 3/2$, and to $I = 0$ and $I = 1$ $Q\bar{Q}qqs$ states 
with $J = 3/2$. This behavior aligns with the 
isospin-independent character of the interaction model described 
in Equation~(\ref{Equ_2}). Nevertheless, this degeneracy is 
nontrivial because of the constraints imposed by the Pauli 
principle.

The dominant quark correlations governed by QCD dynamics,
as summarized in Table~\ref{Tab_2}, allow us to infer 
general features of the 
multiquark spectrum. The mass splitting between the 
spin-triplet and spin-singlet $Q\bar{Q}$ configurations 
corresponds to 
the $J/\psi - \eta_c$ mass difference in the charmonium sector and 
the $\Upsilon - \eta_b$ difference in the bottomonium sector.
Additionally, the mass splitting 
between spin-triplet and spin-singlet $qq$ 
diquarks, with color--flavor configurations 
$\{{\bf \bar{3}}_c, {\bf 6}_f\}$ 
and 
$\{{\bf \bar{3}}_c, {\bf \bar{3}}_f\}$, respectively, 
has been estimated from lattice QCD 
simulations to lie between 
100 and 200~MeV~\cite{Fra21,Ale06,Gre10}.
Based on these findings,
we adopt the following effective values for these mass differences:
\begin{align}\label{mass}
\Delta M^{c\bar c} = M^{c\bar c}_{\{{\bf 1}_c, {\bf 1}_f, 1_s\}} - M^{c\bar c}_{\{{\bf 1}_c, {\bf 1}_f, 0_s\}} &= 86 \,\, {\rm MeV} \nonumber \, , \\
\Delta M^{b\bar b} = M^{b\bar b}_{\{{\bf 1}_c, {\bf 1}_f, 1_s\}} - M^{b\bar b}_{\{{\bf 1}_c, {\bf 1}_f, 0_s\}} &= 61 \,\, {\rm MeV} \nonumber \, , \\
\Delta M^{qq} = M^{qq}_{\{{\bf \bar 3}_c, {\bf 6}_f, 1_s\}}  - M^{qq}_{\{{\bf \bar 3}_c, {\bf \bar 3}_f, 0_s\}} &= 146 \,\, {\rm MeV} \, .
\end{align}

Thus, the mass of a $Q\bar Q qqq$ pentaquark state can be written as
\begin{equation}
M^{Q\bar Q qqq}_i = M_0^{Q \bar Q,q} - B_i + \Delta M^{Q\bar Q} \, \delta_{s_2,1} + \Delta M^{qq} \, \delta_{s_3,1} \, ,
\end{equation}
where $B_i$ is the binding energy given in Table~\ref{Tab_5} and 
$M_0^{Q \bar Q,q}$ denotes the sum of the masses of a spin-zero $Q\bar Q$ 
pair, a spin-zero $qq$ diquark, and a light quark.
A similar expression holds for the $Q\bar{Q}qqs$ states,
where the light quark is replaced by a strange quark, and the 
corresponding mass term is denoted $M_0^{Q \bar Q,s}$:
\begin{equation}
M^{Q\bar Q qqs}_i = M_0^{Q \bar Q,s} - B_i + \Delta M^{Q\bar Q} \, \delta_{s_2,1} + \Delta M^{qq} \, \delta_{s_3,1} \, .
\end{equation}

Using $M_0^{c\bar c,q}=4319$~MeV, we compute the predicted masses listed 
in Table~\ref{Tab_6}. The recent observation of a hidden-charm 
pentaquark with strangeness~\cite{Che22} allows for the calibration of 
the free parameter in the strange sector. Adopting 
$M_0^{c\bar c,s}=4471$~MeV, we obtain the 
results shown in Table~\ref{Tab_8}. For completeness, 
Table~\ref{Tab_9} presents a comparison of our 
predictions with those from other theoretical approaches.
\begin{table}[H]
\caption{{Predicted} properties of $c\bar{c}qqq$ pentaquark 
states compared to experimental data.}
\begin{tabularx}{\textwidth}{cCcCc} 
\toprule
\textbf{Vector}  & \boldmath{$(I)J^P$}     & \boldmath{$M_{\rm Th}$} \textbf{(MeV)} &      \textbf{State}          & \boldmath{$M_{\rm Exp}$} \textbf{(MeV)~\cite{Aai15,Aai19}}                \\ \midrule
$v_1$   & $(1/2)1/2^-$ & 4312               &       $P_c(4312)^+$ & $4311.9 \pm 0.7 ^{+6.8}_{-0.6}$     \\ 
$v_2$   & $(1/2)1/2^-$ & 4388               & {$P_c(4380)^+$} & {$4380 \pm 8 \pm 29$}             \\
$w_1$   & $(1/2)3/2^-$ & 4393               &                                &                       \\
$v_3$   & $(1/2)1/2^-$ & 4441               &       $P_c(4440)^+$ & $4440.3 \pm 1.3 ^{+4.1}_{-4.7}$     \\ 
$w_3$   & $(3/2)3/2^-$ & 4453               &       $P_c(4457)^+$ & $4457.3 \pm 0.6 ^{+4.1}_{-1.7}$    \\  
\bottomrule
\end{tabularx}
\label{Tab_6}
\end{table}
\unskip
\begin{table}[H]
\caption{{Predicted} properties of $c\bar{c}qqs$ pentaquark states
compared to experimental data.}
\begin{tabularx}{\textwidth}{cCcCc} 
\toprule
\textbf{Vector}  & \boldmath{$(I)J^P$}   &   \boldmath{$M_{\rm Th}$} \textbf{(MeV)} &  \textbf{State}                      &  \boldmath{$M_{\rm Exp}$} \textbf{(MeV)~\cite{Che22,Aak21}}  \\ \midrule
$v_1$   & $(0)1/2^-$ & 4338                 &  $P_{\Psi s}^\Lambda(4338)$ &  $4338 \pm 0.7$        \\ 
$v_2$   & $(0)1/2^-$ & 4419                 &                             &                     \\
$w_1$   & $(0)3/2^-$ & 4423                 &                             &               \\
$v_3$   & $(0)1/2^-$ & 4474                 &  $P_{cs}(4459)$             &  $4454.9 \pm 2.7$      \\ 
$w_3$   & $(1)3/2^-$ & 4483                 &                             &  $4467.8 \pm 3.7$      \\  
\bottomrule
\end{tabularx}
\label{Tab_8}
\end{table}
\unskip
\begin{table}[H]
\caption{{Mass} predictions (in MeV) of $c\bar{c}qqs$ pentaquarks
from various models.}
\begin{tabularx}{\textwidth}{cCcCcC} 
\toprule
 \boldmath{$(I)J^P$}  &   \textbf{This work }         &\textbf{ Ref.~\cite{Wul17}} & \textbf{Ref.~\cite{Wag20}}      & \textbf{Ref.~\cite{Hup22}}  & \textbf{Ref.~\cite{Fer20}} \\ \midrule
 $(0)1/2^-$ & 4338                 & 4362.3            & $4319.4^{+2.8}_{-3.0}$ & 4330               & 4474        \\ 
 $(0)1/2^-$ & 4419                 & 4548.2            & $4456.9^{+3.2}_{-3.3}$ & 4475               & 4522            \\
 $(0)3/2^-$ & 4423                 & 4556.1            & $4423.7^{+6.4}_{-6.8}$ & 4440               & 4522          \\
 $(0)1/2^-$ & 4474                 & 4571.4            & $4463.0^{+2.8}_{-3.0}$ & 4476               & $-$          \\ 
 $(1)3/2^-$ & 4483                 & 4846.4            & $-$                    & $-$                & $-$      \\  
\bottomrule
\end{tabularx}
\label{Tab_9}
\end{table}
\vspace{-6pt}

We are now in a position to make parameter-free predictions 
for the lowest-lying 
nonstrange and strange hidden-bottom pentaquarks, for which no
experimental observations currently exist. These predictions
are presented in 
Tables~\ref{Tab_7} and~\ref{Tab_10}, alongside results 
from other theoretical frameworks. Together, they provide
a consistent set of spin-parity assignments for the 
lightest hidden-flavor pentaquark states and serve as useful 
benchmarks for future experimental investigations, particularly in the 
hidden-bottom sector.
\begin{table}[H]
\caption{Mass predictions (in MeV) of nonstrange $b\bar{b}qqq$ pentaquarks from various models.}
\begin{tabularx}{\textwidth}{CCCCC} 
\toprule
 \boldmath{$(I)J^P$ }    & \textbf{This Work}          & \textbf{Ref.~\cite{Wul17}} & \textbf{Ref.~\cite{Yan19}} & \textbf{Ref.~\cite{Fer19}}  \\ \midrule
 $(1/2)1/2^-$ & {11062       } 
       & 11137.1           & 11080 (11078)     & 10605  \\ 
 $(1/2)1/2^-$ & 11121              & 11148.9           & 11115 (11043)     & 10629  \\
 $(1/2)3/2^-$ & 11122              & 11237.5           & 11124 (11122)     & 10629  \\
 $(1/2)1/2^-$ & 11195              & 11205.0           & $-$               & $-$ \\ 
 $(3/2)3/2^-$ & 11207              & 11370.6           & 11112 (10999)     & $-$  \\  
\bottomrule
\end{tabularx}
\label{Tab_7}
\end{table}
\unskip
\begin{table}[H]
\caption{{Mass} predictions (in MeV) of $b\bar{b}qqs$ pentaquarks from 
various models.}
\begin{tabularx}{\textwidth}{CCCC} 
\toprule
 \boldmath{$(I)J^P$}   & \textbf{This Work }         & \textbf{Ref.~\cite{Wul17}} & \textbf{Ref.~\cite{Fer20}}    \\ \midrule
 $(0)1/2^-$ & 11088              & 11117.7           & 10671   \\ 
 $(0)1/2^-$ & 11147              & 11183.8           & 10695   \\
 $(0)3/2^-$ & 11148              & 11180.2           & 10695   \\
 $(0)1/2^-$ & 11224              & 11301.2           & $-$     \\ 
 $(1)3/2^-$ & 11233              & 11509.0           & $-$     \\  
\bottomrule
\end{tabularx}
\label{Tab_10}
\end{table}
\vspace{-6pt}

The spin-parity quantum numbers of hidden-flavor pentaquarks 
remain experimentally undetermined~\cite{Bra20}. Nevertheless, a 
range of theoretical models offer predictions that can be confronted 
with our results summarized in Tables~\ref{Tab_6}--\ref{Tab_10}. 
We begin our discussion with the hidden-charm sector
for subsequent analysis of the remaining systems.

Although the quantum numbers of the $P_c(4312)^+$ remain 
uncertain~\cite{Zhu16}, the majority 
of theoretical works favor a $J^P = 1/2^-$ 
assignment~\cite{Wun10,Wan11,Yan12,Wul12,Xia13,Yaa17}. 
The two narrow overlapping structures $P_c(4440)^+$ and 
$P_c(4457)^+$~\cite{Aai19} were initially 
interpreted as a single resonance, 
$P_c(4450)^+$~\cite{Aai15}. Before the experimental resolution of this 
doublet, theoretical models had already predicted the existence 
of two nearly degenerate states with quantum numbers 
$J^P = 1/2^-$ and $3/2^-$ in this mass region. These states were 
interpreted either as hidden-charm pentaquarks 
dynamically generated through 
$\Sigma_c \bar D^*$ interactions~\cite{Wun10,Wum11,Xia13} or as bound 
states of charmonium $\Psi(2S)$ and the nucleon~\cite{Eid16}. In both 
scenarios, the predicted quantum numbers are consistent with our findings.

Our model also predicts two candidate states corresponding to 
the broad $P_c(4380)^+$ 
resonance: one with $J=1/2$ and another with $J=3/2$. The true nature of 
this state remains uncertain and represents a significant 
challenge for future 
experimental investigations~\cite{Kar15}. 
While earlier studies proposed various spin
assignments, ranging from $J=3/2$ to \mbox{$J=5/2$~\cite{Aai15}}, a more 
recent analysis of $B_s \to J/\psi p \bar p$ decays favors a 
$J^P = 3/2^-$ assignment~\cite{Wag21}. Based on this evidence, 
we tentatively associate the $P_c(4380)^+$ with the $J^P = 3/2^-$ 
state predicted by our model. This identification naturally
suggests the existence of a nearby $J^P = 1/2^-$ 
partner state, with an estimated mass around 4390~MeV.
It is worth noting that for strange hidden-charm pentaquarks,
our results are in good agreement with the recent experimental 
measurement by Belle~\cite{Ada25}, which reports a mass of 
$4471.7 \pm 4.8 \pm 0.6$ MeV/c$^2$ for the 
$P_{cs}(4459)$. Our model predicts a mass 4474 MeV/c$^2$
for the $J^P=1/2^-$ state. This consistency is particularly encouraging.

Ref.~\cite{Hua16} investigates hidden-charm pentaquarks 
within the framework of a 
quark delocalization color screening model. 
The most deeply bound configurations, corresponding to $J=1/2$ and 
$J=3/2$, are dominated by a $(Q\bar Q)(qqq)$ structure, reflecting 
strong short-range Coulomb-like correlations between the heavy quarks. 
In contrast, configurations of the form $(q\bar Q)(Qqq)$ result 
only in quasi-bound states.
Ref.~\cite{Yan17} investigates hidden-charm 
pentaquarks within a chiral quark model. In contrast to 
Ref.~\cite{Hua16}, the dominant configuration found 
here is $(q\bar Q)(Qqq)$. 
This difference may be attributed to the absence of shared light quarks 
between the quarkonium and baryon components 
in the $(Q\bar Q)(qqq)$ configuration, which 
suppresses meson exchanges due to the 
Okubo--Zweig--Iizuka (OZI) rule~\cite{Bro90}. This 
rationale also underlies the tendency of hadronic molecular models 
to describe 
hidden-flavor pentaquarks as bound states of open-flavor 
hadrons~\cite{Hos16,Guo18,Xia19,She19,Wan20,Bur15}. 
In such models, $D$-meson 
exchange is too short-ranged to effectively compete 
with the medium-range attraction 
generated by light-meson exchanges, which is more naturally
realized in the $(q\bar Q)(Qqq)$ configuration.
Consequently, hybrid and purely gluonic models tend to 
yield different multiquark spectra.
Several studies also predict the existence of $I=3/2$ pentaquarks. 
Within the hadroquarkonium framework, Ref.~\cite{Per16} 
provides robust predictions of 
isospin-3/2 $\Psi(2S)-\Delta$ bound states near 4.5~GeV. From the 
constituent quark model perspective, Ref.~\cite{Ric17} 
similarly supports the existence of $I=3/2$ hidden-flavor pentaquarks, 
including a $J=5/2$ state located close to threshold. 
Ref.~\cite{Wul17} explores compact hidden-flavor pentaquarks constructed 
from the repulsive color-octet--octet 
configuration 
${\bf 8}{(Q\bar Q)} \otimes {\bf 8}{(qqq)}$. 
In this framework, negative-parity states are 
predicted to remain bound only with respect to the heavier 
$(q\bar Q)(Qqq)$ threshold. A distinctive feature of this model is the 
strong suppression of branching ratios into hidden-flavor channels,
which serves to differentiate 
compact color-octet pentaquarks from meson--baryon molecular states.
Ref.~\cite{Wen19} employs an extended chromomagnetic model 
that augments the conventional color--spin interaction with 
effective quark-pair 
mass parameters, fitted to reproduce known
meson and baryon spectra. The $I=1/2$ spectrum, 
as presented in {Figure 1 } 
of Ref.~\cite{Wen19}, 
aligns well with our findings. 
However, it does not reproduce the near-degeneracy between $I=1/2$ and 
$I=3/2$ states observed in our model, likely a consequence
of the method used to extract the effective 
parameters, since the underlying interaction is isospin 
independent.

Beyond constituent quark models, a variety 
of other theoretical frameworks 
have been employed to study hidden-flavor pentaquarks. 
Predictions from diquark-based models exhibit significant variation,
largely depending on the assumptions made regarding 
diquark dynamics~\cite{Mai15,Gir19,Ali19,Shi21,Mai14}. 
QCD sum rule approaches also lead to diverse outcomes, 
predicting both 
molecular-like states~\cite{Zha19,Che19,Azi17,Wan21} and hidden-color 
configurations~\cite{Pim20}. For a thorough overview of recent 
progress in QCD sum rule techniques and their relevance 
to exotic hadrons, see Ref.~\cite{Nar21}.
Hadronic molecular models based on effective chiral Lagrangians or 
one-boson exchange potentials depend on low-energy constants 
and couplings that are often poorly known. These parameters are 
typically estimated using quark model 
relations~\cite{Wun10,Yam17,Men19,Wan19,Yal21}.

\clearpage 

For the hidden-bottom pentaquarks listed in Table~\ref{Tab_7}, 
only theoretical predictions are available for comparison. 
Ref.~\cite{Wul17} employs a color-magnetic interaction model,
estimating mass splittings relative to a reference mass fitted to 
experimental data. Ref.~\cite{Yan19} uses a chiral quark model and 
solves the five-body bound-state problem via the Gaussian expansion 
method. We cite results from the color-singlet calculation, which is 
theoretically closest to our framework, and list in parentheses those 
from the coupled-channel calculation that includes hidden-color 
configurations. Ref.~\cite{Fer19} provides predictions from a 
hadroquarkonium model for isospin-$1/2$ baryons, 
considering two scenarios with different 
chromoelectric polarizabilities. We report the results 
corresponding to the weaker interaction strength, in which
hidden-bottom pentaquarks appear in the 10.6--10.9~GeV range. 
The alternative model, based on 
a purely Coulombic treatment of charmonia, predicts 
more deeply bound states in the 10.4--10.7 GeV range. 
In this case, positive-parity 
states are less bound and lie approximately 150~MeV above their 
negative-parity counterparts.
It is noteworthy that quark-based approaches~\cite{Wul17,Yan19}
place hidden-bottom pentaquarks in 
the 11.0--11.2~GeV mass range, whereas the 
hadroquarkonium model~\cite{Fer19}
predicts more deeply bound states.

In the strange hidden-charm sector, the perturbative color-magnetic 
model of Ref.~\cite{Wul17} predicts large mass splittings among the 
lowest-lying states. 
In contrast, the chiral effective field theory potentials 
of Ref.~\cite{Wag20} produce 
results more consistent with ours, locating the lowest 
states in the 4.3--4.4~GeV region. Notable, these authors 
find no bound states in 
$I=1$ channels. Ref.~\cite{Hup22} employs a chiral quark model 
combined with a variational method, expanding radial wave functions
in Gaussian bases. This approach, which has proven effective in 
describing nonstrange pentaquarks, 
predicts small binding energies across various baryon--meson channels.
A complementary study of $c\bar c qqs$ tetraquarks 
within a similar chiral quark framework,
utilizing the Gaussian expansion approach in combination with the 
complex-scaling technique, is presented in Ref.~\cite{Yan24}.
By contrast, the hadroquarkonium model of 
Ref.~\cite{Fer20} predicts higher masses for strange 
hidden-charm pentaquarks while yielding lower masses 
than quark-based models for nonstrange and strange 
hidden-bottom systems.

The predicted strange hidden-bottom pentaquark states lie 
close in mass to their nonstrange 
counterparts, within the 11.1--11.3~GeV range. This aligns 
with the perturbative chromomagnetic model of 
Ref.~\cite{Wul17}. As in the nonstrange sector, the 
hadroquarkonium model of Ref.~\cite{Fer20} predicts 
significantly lower masses, with 
the most deeply bound states appearing near 10.4~GeV under 
the strong-attraction scenario. Consequently, 
hidden-bottom pentaquarks, both strange 
and nonstrange, offer a valuable testing ground for distinguishing 
between different multiquark interaction mechanisms. 
In particular, models based on quark 
substructure suggest that experimental searches should 
concentrate around the 11~GeV region.

Alternative theoretical frameworks have also been 
employed to study both strange and 
nonstrange hidden-flavor pentaquarks. One such approach is the 
diquark--triquark model of Ref.~\cite{Zhu16}, 
which leverages attractive color 
configurations involving non-pointlike triquarks and uses a light-cone 
distribution amplitude to describe the internal structure of
the pentaquark. The model incorporates an 
effective Hamiltonian with spin--orbit interactions and predicts,
among other states, 
the $P_c(4380)^+$ as a $J^P=3/2^-$ resonance with a mass of 4349~MeV and 
the originally reported $P_c(4450)^+$~\cite{Gar22} as a $J^P=5/2^+$ state 
with a mass of 4453~MeV. The lightest 
predicted state is a $J^P=3/2^-$ hidden-charm 
pentaquark at 4085~MeV. Although the model discusses 
$J^P=1/2^-$, their masses are not explicitly provided. 
Extending this framework, the same authors also explore
strange hidden-charm and hidden-bottom pentaquarks. 
The lightest states in each sector for each spin-parity 
assignment are
\vspace{-6pt}
$$
\left\{
\begin{array}{ll}
\text{Hidden-charm:} & (3/2^-, 5/2^-, 5/2^+) = (4085, 4433, 4453) 
\text{ MeV} \\ \\
\text{Strange hidden-charm:} & (3/2^-, 5/2^-, 5/2^+) = 
(4314, 4624, 4682) \text{ MeV}  \\ \\
\text{Hidden-bottom:} & (3/2^-, 5/2^-, 5/2^+) = 
(10723, 11045, 1146) \text{ MeV}  \\ \\
\text{Strange hidden-bottom:} & (3/2^-, 5/2^-, 5/2^+) = 
(10981, 11264, 11413) \text{ MeV}\, .
\end{array}
\right.
$$

Preliminary analyses of experimental data in the hidden-charm sector 
suggest the coexistence of both negative- and positive-parity states 
within the same energy region~\cite{Aai15}. We 
calculated the mass of the lowest positive-parity state, 
which corresponds to two nearly degenerate configurations 
with quantum numbers $J^P=1/2^+$ and $3/2^+$.
This excitation arises from the most strongly 
correlated internal structure, featuring a
$Q\bar Q$ pair in the color--flavor--spin configuration
$\{{\bf 1}_c, {\bf 1}_f, 0_s\}$, and a $qq$ subsystem in 
$\{{\bf \bar 3}_c, {\bf \bar 3}_f, 0_s\}$.
The lightest positive-parity hidden-charm pentaquark states 
are found above 4.5~GeV, specifically 4527~MeV for $c\bar c udd$ 
and 4552~MeV for $c\bar c uds$. In contrast, 
excitation energies in the hidden-bottom sector
are significantly lower, resulting in positive-parity 
states that lie closer in mass to their negative-parity 
ground-state counterparts:
11,273~MeV for $b\bar b udd$ and 
11,291~MeV for $b\bar b uds$. Nearly degenerate 
opposite-parity states may originate from distinct internal 
mechanisms, for example, a negative-parity state formed
by coupling a radially excited quarkonium 
to a nucleon versus a positive-parity state generated 
through orbital excitation of a ground-state 
quarkonium--nucleon system~\cite{Eid16}. This pattern of parity 
assignments is also consistent with the Born--Oppenheimer 
interpretation proposed in Ref.~\cite{Gir19}. 

The decay properties of multiquark states are highly sensitive to their 
internal \linebreak structure~\cite{Wul17}. Transitions into heavy meson--heavy 
baryon channels are generally expected to be strongly suppressed,
primarily due to 
the need for heavy meson exchange in $t$-channel processes. Given 
their quark composition, these exotic states may decay in a manner
analogous to excited quarkonia such as 
$\Psi(nS)$, $\Upsilon(nS)$, $\eta_c(nS)$, or 
$\eta_b(nS)$. As a result, configurations containing a spin-zero 
$Q\bar{Q}$ pair (e.g., $v_1$, $v_3$, $w_3$ in Table~\ref{Tab_2}) are 
predicted to exhibit narrower decay widths
than those containing a spin-one 
$Q\bar{Q}$ pair (such as $v_2$, $w_1$). This pattern is consistent
with current 
empirical findings (see Table~\ref{Tab_1}). Nonetheless, the total 
decay width also depends critically on the internal structure 
of the bound state~\cite{Gar18,Gai22}, 
a topic discussed in detail in Section~\ref{secIV}.

It should be emphasized that the correlation mechanisms employed in our 
framework do not universally guarantee 
the stability of multiquark systems 
across all flavor configurations. Just as the short-range repulsion 
in the nucleon--nucleon interaction, originating from one-gluon
exchange, 
is not a general feature of all hadronic systems, 
our QCD-based model does not impose 
constraints on the color configurations of pentaquarks 
containing an anticharm 
or antibottom quark, such as $\bar{Q}qqqq$. As discussed in 
Ref.~\cite{Ric19}, these states are unlikely to be bound due to the 
competing effects of chromoelectric and chromomagnetic interactions, 
which tend to disfavor binding in such configurations.

Our findings also contribute to the ongoing discussion regarding
the possibility of 
charmonium forming bound states with atomic nuclei, a scenario 
originally proposed by Brodsky~\cite{Bro90}. Since charmonium contains 
no valence $u$ and $d$ quarks, its interaction with nucleons via light 
meson exchange is suppressed by the OZI rule. This limitation has 
led to the development of alternative models to generate 
sufficient attraction, including those based on
chromoelectric polarizability and the energy--momentum tensor 
distributions of the nucleon~\cite{Eid16,Per16,Dub08}. 
Additional support 
comes from studies of in-medium modifications of hadronic 
structure, such as the predicted formation
of $J/\psi$-nucleus bound states~\cite{Tsu11}, and 
similar theoretical expectations for $\eta_c$-nucleus
systems across various nuclear 
targets~\cite{Cob20}. Our approach introduces 
a distinct mechanism, driven by 
short-range one-gluon exchange between the charmonium and 
the constituent quarks of the nucleon. 
This mechanism, originally proposed in the context of 
dibaryon resonances~\cite{Oka80,Gol89,Pan01,Val01,Val05}, represents, to 
our knowledge, the first derivation of such bound-state behavior solely 
from quark--gluon dynamics within a truncated Hilbert space.

\section{Molecules of Heavy Hadrons}
\label{secIII}
A striking prediction of the quark model
was the possible existence of doubly heavy 
tetraquarks, $QQ\bar q\bar q$~\cite{Ade82}.
Their stability was shown to depend on the mass ratio $M_Q/m_q$. 
The mechanism that stabilizes such states at large $M_Q/m_q$ 
is analogous to that which makes the hydrogen molecule 
significantly more stable than 
positronium in atomic physics~\cite{Ric93}. 
This prediction was confirmed by the LHCb Collaboration,
which reported the 
discovery of a very narrow peak in the $DD\pi$ spectrum,
identified as the 
$T_{cc}^+$ state~\cite{Aai22,Aaj22}. 
It corresponds to a minimal quark content 
$cc\bar u\bar d$. 
The $T_{cc}^+$ lies just above the strong interaction threshold, 
missing binding by a very small amount. 
Consequently, it is widely believed that its heavier counterparts, 
$T_{bc}=bc \bar u\bar d$ and $T_{bb}=bb\bar u\bar d$, could be stable 
against both strong and electromagnetic decays~\cite{Aai22,Col24}.
Theoretical estimates of the binding energy of the $T_{bb}$ tetraquark
vary, with reported values ranging from 90 to 
214 MeV~\cite{Ade82,Ric18,Eic17,Vij09,Fra17,Kar17,Bic16,Luo17,Duc13,Cza18,Jun19}.

As shown in Ref.~\cite{Ric18}, the probability of 
the color $\bf 6 \bf \bar 6$ component in a $QQ\bar q\bar q$ 
tetraquark tends to zero as $M_Q \to \infty$. Consequently,
doubly heavy tetraquarks are expected to approach 
a pure $\bf{\bar 3} \bf{3}$ color configuration rather
than forming a single color-singlet meson--meson $\bf 1 \bf 1$ 
molecular state~\cite{Her20}. These two-body color components
can be expanded into a mixture of several physical 
meson--meson channels~\cite{Har81}. For instance, in the case
of the isoscalar axial-vector $bb\bar u \bar d$ 
tetraquark, the relevant channels are
$BB^*$ and $B^*B^*$, see Table II of Ref.~\cite{Via09}.
As a result, $QQ\bar q\bar q$ tetraquarks are naturally
generated through a coupled-channel system of colorless 
meson--meson states~\cite{Car12,Ike14}.

Another area that has recently garnered significant attention 
is the study of heavy baryon bound states,
particularly motivated by advances in lattice QCD.
The HAL QCD Collaboration~\cite{Gon18} investigated the 
$\Omega \Omega$ system in the $^1S_0$ channel, finding 
an overall attractive interaction and reporting a binding energy of
$B^{\rm QCD}_{\Omega\Omega}=1.6(6)\left(^{+0.7}_{-0.6}\right)$~MeV.
When Coulomb repulsion is included, the binding energy 
is reduced by approximately half, yielding
$B^{\rm QCD+Coulomb}_{\Omega\Omega}=0.75(5)(5)$~MeV.
The $\Omega_{ccc}\Omega_{ccc}$ system was also studied
using the HAL QCD method~\cite{Lyu21}. The results
show weak short-range repulsion surrounded by
a relatively strong attractive potential well, resulting in a 
bound state with a binding energy of 
$B^{\rm QCD}_{\Omega_{ccc}\Omega_{ccc}}=5.68(0.77)\left(^{+0.46}_{-1.02}\right)$~MeV.
When Coulomb interaction is taken into account, 
a scattering length of
$a^C_0 = -19(7)\left(^{+7}_{-6}\right)$~fm is obtained.
More recently, a lattice QCD study
of the $\Omega_{bbb}\Omega_{bbb}$ system
in the $^1S_0$ channel~\cite{Mat23} reported 
a deeply bound state with a binding energy of
$B^{\rm QCD}_{\Omega_{bbb}\Omega_{bbb}}= 81\left(^{+16}_{-14}\right)$~MeV.
At this depth, Coulomb repulsion acts merely as a perturbative correction,
reducing the binding energy by approximately 5 and 10~MeV. 

The possible existence of deuteron-like hadronic molecular states 
composed of vector--vector or pseudoscalar--vector meson pairs
was first proposed in Ref.~\cite{Tor91}. This framework, in which
meson--meson systems are stabilized through some form
of interhadron potential, has since become a common approach
to interpreting potential hadronic
\linebreak molecules~\cite{Man93,Eri93,Clo10} (see
Ref.~\cite{Che16} for a recent review).
Moreover, insights from nuclear 
physics indicate that if an attractive interaction exists
between two baryons, the presence of additional nucleons,
provided no severe Pauli principle constraints are introduced,
can enhance the overall binding.
Simple examples highlight this effect:
the deuteron, with quantum numbers $(I)J^P=(0)1^+$, 
has a binding energy of $2.225$~MeV; the triton,
$(I)J^P=(1/2)1/2^+$, is bound by $8.480$~MeV; 
and the $\alpha$ particle, $(I)J^P=(0)0^+$,
has a binding energy of $28.295$~MeV. Consequently, 
the binding energy per nucleon, $B/A$, increases 
roughly in the ratio 1:3:7.
A comparable trend is observed in systems with strangeness $-1$.
The hypertriton $^3_\Lambda$H, with $(I)J^P=(0)1/2^+$, is 
weakly bound with a separation
energy of $130 \pm 50$~keV~\cite{Jur73}, whereas
the $^4_\Lambda$H, $(I)J^P=(0)0^+$, has a significantly higher
separation energy of 
$2.12 \pm 0.01 \, {\rm (stat)} \, \pm 0.09 \, {\rm (syst)}$~MeV~\cite{Ess15}. 

Thus, an important and challenging question is whether the existence
of deeply bound two-hadron systems, like the $T_{bb}$ tetraquark
or the $\Omega_{bbb}\Omega_{bbb}$ dibaryon predicted by lattice 
QCD, could lead to the formation of bound states in
systems involving more than two 
hadrons~\cite{Gac18,Gar17,Mam19,Mar08,Mar20,Wul23,Gar24}.
This issue is particularly
relevant in the bottom sector, where the predicted 
binding energies for two-body systems are exceptionally 
large. In the following, we analyze the
three-meson and three-baryon cases separately.

\subsection{The Three-Meson System}
\label{subsecIII.II} 
We solve the Faddeev equations for the three-meson bound-state problem, using as input the 
two-body $t-$matrices from Refs.~\cite{Vij09,Via09}. These
interactions give rise to an isoscalar axial-vector 
$bb\bar u \bar d$ bound state, 
the $T_{bb}$, which emerges from a coupled-channel
system involving both pseudoscalar--vector and vector--vector 
combinations of two $B$~mesons.
The three-body channels most conducive to generating bound states 
are those that include the $T_{bb}$ tetraquark and exclude
two-$B$-meson components, due to the lack of attraction in 
the $BB$ interaction~\cite{Ade82,Ric18,Eic17,Vij09,Fra17,Kar17,Bic16,Luo17,Duc13,Cza18,Jun19}. 
Channels with total angular momentum $J=0$ or $1$ would necessarily 
involve $BB$ subsystems and are therefore disfavored. 
In contrast, channels with $J=3$ do not 
contain two-body subsystems with $j=1$, which are required for the 
$T_{bb}$ quantum numbers, making them unsuitable as well. 
A similar exclusion applies to channels 
with isospin $I=3/2$. Consequently, the three-body channel with quantum 
numbers $(I)J^P = (1/2)2^-$ is the only configuration
that satisfies all the 
conditions necessary to maximize the likelihood of binding. The two-body 
channels contributing to this state are listed 
in Table~\ref{Tab_12}.

\begin{table}[H]
\caption{{Different} 
two-body channels $(i,j)$ contributing to the $(I)J^P=(1/2)2^-$ configuration of the
$B B^* B^* - B^* B^* B^*$ three-body system.}
\begin{tabularx}{\textwidth}{CCC} 
\toprule
\textbf{Interacting Pair} & \boldmath{$(i,j)$} & \textbf{Spectator}  \\ 
\midrule
\multirow{2}{*}{$B B^*$}   & $(0,1)$   & {$B^*$} \\
                           & $(1,1)$   &                      \\
\multirow{2}{*}{$B^* B^*$} & $(0,1)$   & {$B^*$} \\
                           & $(1,2)$   &                      \\
$B^* B^*$                  & $(1,2)$   & $B$ \\ \bottomrule																																				
\end{tabularx}
\label{Tab_12}
\end{table}
\vspace{-6pt}

\begin{figure}[H]
\includegraphics[width=.7\columnwidth]{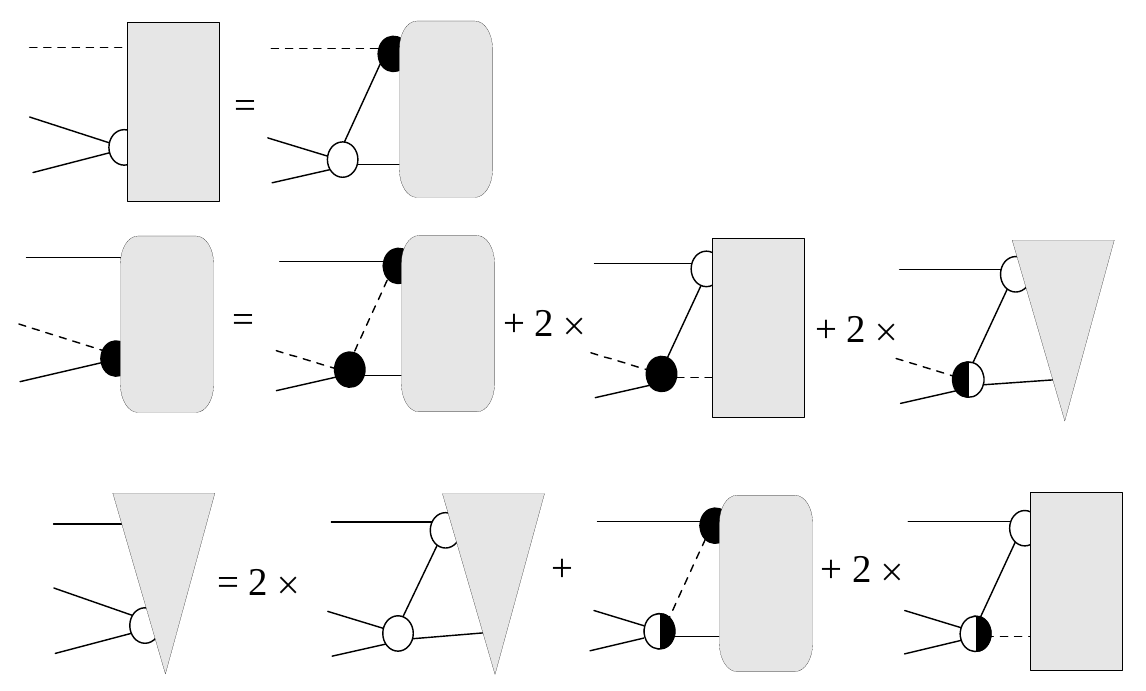}
\caption{{Diagrammatic} 
representation of the Faddeev equations for the three-$B$-meson system.}
\label{Fig_2}
\end{figure}

The Faddeev equations for the bound-state three-meson 
problem~\cite{Fad61,Fad65} can be written~as
\begin{equation}
T_i = t_i G _0(T_j + T_k) \, ,   
\label{ER_5}
\end{equation}
with
\begin{equation}
t_i = V_i + V_i G_0 t_i \, ,
\label{ER_6}
\end{equation}
where $t_i$ denotes the two-body $t$-matrices, which already account 
for the coupling among all two-body channels that contribute to a given 
three-body state. For coupled three-meson systems such 
as $BB^*B^*$ and $B^*B^*B^*$, the resulting 
equations are illustrated in Figure~\ref{Fig_2}. In these diagrams, solid 
lines represent $B^*$ mesons and dashed lines correspond to $B$ mesons.

If the last term on the right-hand side of the second equation 
in Figure~\ref{Fig_2} is omitted, the first and second equations 
reduce to the Faddeev equations for a system consisting of two identical 
bosons and a third distinguishable particle~\cite{Gar17}. Likewise, by
omitting the last two terms in the third equation, one recovers 
the Faddeev equation for a system of three identical bosons, 
as in this case all three coupled Faddeev equations become 
identical~\cite{Gar17}. The 
additional terms shown in Figure~\ref{Fig_2} are, naturally, 
those responsible 
for the coupling between the $BB^*B^*$ and $B^*B^*B^*$ components.
\vspace{-9pt}

\begin{figure}[H]
\includegraphics[width=0.6\columnwidth]{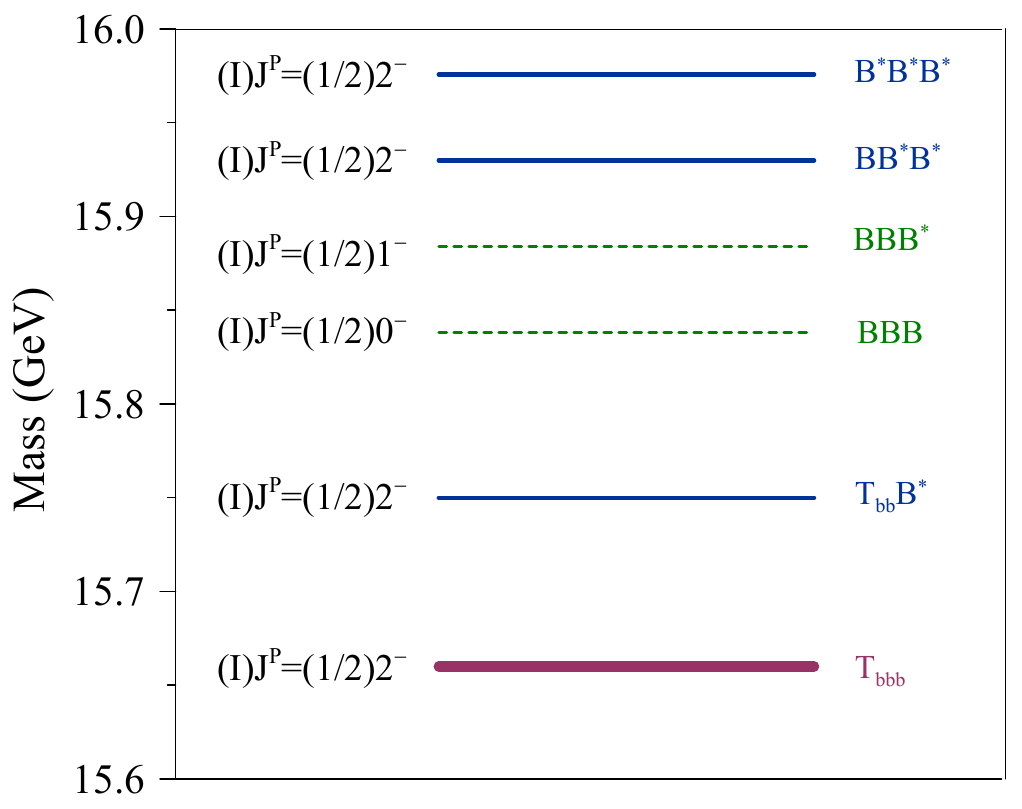}
\caption{Mass of the three-body $BB^*B^*{-}B^*B^*B^*$ bound state, 
with quantum numbers \mbox{$(I)J^P = (1/2)2^-$} and
denoted as $T_{bbb}$ (represented by the 
thick purple line), shown in comparison with strong
decay thresholds (solid blue lines) 
and electromagnetic decay thresholds (dashed green lines).}
\label{Fig_3}
\end{figure}

Figure~\ref{Fig_3} displays the results of our calculation. The solid 
blue lines correspond to the strong decay thresholds of the 
three-$B$-meson system $BB^*B^*{-}B^*B^*B^*$ with quantum numbers 
$(I)J^P = (1/2)2^-$. These thresholds include 
the $B^*B^*B^*$, $BB^*B^*$, and $T_{bb}B^*$ channels. Dashed green 
lines indicate electromagnetic decay thresholds involving three 
$B$ mesons, specifically the $BBB^*$ and $BBB$ channels 
with quantum numbers $(I)J^P = (1/2)1^-$ 
and $(1/2)0^-$, respectively. The thick purple line represents the 
energy of the $T_{bbb}$ state, located 90~MeV below the lowest decay 
threshold. These results are derived using the binding 
energy of the axial-vector $T_{bb}$ tetraquark reported 
in Ref.~\cite{Fra17}. 

There also exists a baryon--antibaryon threshold, 
$\Omega_{bbb} \, \bar{p}$, 
which is clearly decoupled from the $T_{bbb}$ state. This 
decoupling arises from the 
dominant tetraquark--meson structure of $T_{bbb}$ and the orthogonality 
of its color wave functions to that of the baryon--antibaryon system. 
Based on the $T_{bb}$ binding energy 
reported in Ref.~\cite{Fra17}, the mass 
of the $\Omega_{bbb}$ baryon is estimated to be 14.84~GeV, placing the 
$\Omega_{bbb} \, \bar{p}$ threshold at approximately 15.78~GeV, 
well above the mass of the $T_{bbb}$ 
state. Although theoretical predictions of 
the $\Omega_{bbb}$ mass vary considerably
(see Table~1 of Ref.~\cite{Zha17}),
even if the $\Omega_{bbb} \, \bar{p}$ 
threshold were to lie below the $T_{bbb}$ mass, 
the latter would still manifest as a narrow resonance.
This is because $\left| \Psi_{T_{bbb}} \right\rangle$
is an eigenstate of the Hamiltonian $H$, and the transition 
operator $T$, approximated by $e^{iH}$, cannot induce a
decay into a baryon ($B_1$) and an 
antibaryon ($\bar{B}_2$) if the matrix element 
$\left\langle B_1 \bar{B}_2 | T | \Psi_{T_{bbb}} \right\rangle$
vanishes. This occurs when the overlap
$\left\langle B_1 \bar{B}_2 | \Psi_{T_{bbb}} \right\rangle$ is 
zero~\cite{Lea89}.
The formation of the $T_{bbb}$ state is driven by the attractive 
dynamics of
the three-meson system, whereas the weakly interacting baryon--antibaryon
channel would primarily serve as a detection mode. 
As we will discuss in Section~\ref{secIV}, 
this situation is analogous to the case of 
the $P^+_c(4380)$ pentaquark, which decays 
into the $J/\psi \, p$ channel with a 
width of $\Gamma = 205 \pm 18 \pm 86$ MeV despite having 
a phase space of 
about 345~MeV.

We estimated the binding energy of the three-body 
$T_{bbb}$ system over a range of input binding energies
for the $T_{bb}$ tetraquark, starting from the lowest 
reported value of approximately 90~MeV, as cited 
in Ref.~\cite{Bic16}. Our calculations indicate 
that the $T_{bbb}$ three-meson bound state remains robustly 
stable throughout the entire range considered. 
The binding energy of $T_{bbb}$ decreases from
90~MeV to 43~MeV as the $T_{bb}$ binding energy is varied
between 180~MeV and 87~MeV.~{{The} 
binding energy of the
$T_{bbb}$ system 
is calculated relative to the lowest strong decay threshold, defined as 
$m_B + 2 m_{B^*} - B(T_{bb})$.}
However, if the $T_{bb}$ binding energy is reduced further to 
around 50~MeV, the $T_{bbb}$
system would remain bound by only about $\sim$23~MeV, placing it 
approximately 19~MeV above the lowest $BBB$ threshold. 

The situation is notably less 
favorable in the charm sector, primarily 
due to the larger mass difference between vector and pseudoscalar 
mesons: 141~MeV in the charm sector compared to only 45~MeV in 
the bottom sector. As a consequence, the $DDD$ and $DDD^*$ thresholds 
lie 282~MeV and 141~MeV, respectively, below the $DD^*D^*$ configuration.
In other words, when the internal two-body thresholds of a three-body 
system are widely separated, they tend to destabilize the overall system. 
This is a well-known phenomenon observed in studies of three-body systems 
comprising either two identical baryons or two identical mesons 
along with a 
bound two-body subsystem. Even when strong attractive forces exist 
within some subsystems, the full three-body configurations 
often remain unbound and may even display a pronounced repulsive 
character~\cite{Gar17}.

Finally, it is worth noting that several studies in the literature 
have been developed within theoretical frameworks that are not
directly related to the consistent approach adopted in this
review. For instance, Ref.~\cite{Wul19} predicts the existence of
$DDK$ and $DDDK$ molecules under the assumption 
that the $DK$ interaction is sufficiently strong to generate
a bound state, identified with
the $D^*_{s0}(2317)$. Likewise, Ref.~\cite{Ort24} predicts a $D^*D^*D^*$
bound state based on the hypothesis that the $T_{cc}^+$ is 
itself a $D^*D^*$
bound state, thereby neglecting contributions from $DD^*$ channels
and the corresponding thresholds. Moreover, no mention is made 
of baryon--antibaryon thresholds in that work.

\subsection{The Three-Baryon System}
\label{subsecIII.III}

A situation analogous to that described for the 
three-meson system also arises in the context of the
three-baryon systems. 
The key distinction, however, is that in the baryonic case, the 
binding of the two-body subsystems does not originate from 
coupled-channel dynamics. Given the bound-state nature of 
systems composed of two--$\Omega$ baryons,
predicted across different heavy-flavor sectors 
by lattice QCD calculations~\cite{Gon18,Lyu21,Mat23},
a natural question emerges: could systems 
containing more than two such 
$\Omega$-like baryons also form 
bound states~\cite{Wul23,Gar24,Gar25}? {{We} 
will 
henceforth use the notation
$\Omega^i$ to refer to the $\Omega$ 
for $i=s$, the $\Omega_{ccc}$ for $i=c$, 
and the $\Omega_{bbb}$ for $i=b$.}

Owing to Fermi--Dirac statistics, the $\Omega^i\Omega^i$ system is
restricted to two $L=0$ partial waves: the spin-singlet $^1S_0$ 
and the spin-quintet $^5S_2$. 
Moreover, there exists only one fully antisymmetric 
three-body $\Omega^i\Omega^i\Omega^i$ state 
with all three baryons in relative 
$S$-wave: the \mbox{$J^P = 3/2^+$} state.
In terms of an arbitrary Jacobi coordinate system, denoted as 
$\left|s_{12}, s_3; S \right\rangle$, the fully antisymmetric 
$J^P = 3/2^+$ wave function includes two spin components 
corresponding to $s_{12} = 0$ and $s_{12} = 2$. 
These components must be recoupled using alternative 
Jacobi bases, such as 
$\left|s_1, s_{23}; S \right\rangle$, to properly
account for particle exchange symmetry. The relevant recoupling 
coefficients between different Jacobi coordinate systems are 
provided in Table~\ref{tab1}.

\vspace{-3pt}
\begin{table}[H]
\caption{{Recoupling} 
coefficients between different Jacobi 
coordinate systems for the fully antisymmetric
$\Omega^i\Omega^i\Omega^i$ state with 
quantum numbers $J^P=3/2^+$ state.}
\begin{tabularx}{\textwidth}{CCCC}
\toprule
&&  \multicolumn{2}{c}{$\left|s_{12},s_3;S\right>$}       \\
  && $s_{12}=0$   &  $s_{12}=2$ \\ 
  \midrule
\multirow{2}{*}{$\left|s_1,s_{23};S\right>$}	&   $s_{23}=0$  &   $-\frac{1}{4}$ & $-\frac{\sqrt{5}}{4}$   \\
         &   $s_{23}=2$  & $-\frac{\sqrt{5}}{4}$ & $\frac{3}{4}$  \\ \bottomrule
\end{tabularx}

\label{tab1}
\end{table}
\vspace{-3pt}

This table highlights two key features of the 
$\Omega^i\Omega^i\Omega^i$ state with $J^P=3/2^+$.
First, it demonstrates the dominant contribution of the 
$^5S_2$ partial wave, which significantly outweighs
that of the attractive $^1S_0$ channel.
Second, it reveals the strong mixing
between the $^1S_0$ and $^5S_2$ components within the 
fully antisymmetric three-body wave function.

We solved the three-body $\Omega^i\Omega^i\Omega^i$ system using 
the Faddeev formalism, as described in 
detail in Ref.~\cite{Gar97}. For the two-body input potentials in 
the $^1S_0$ $\Omega^i\Omega^i$, we employed the parametrization from 
Refs.~\cite{Gon18,Lyu21} for the cases $i = s$ and $c$. 
For the $i = b$ sector, we 
parametrized the potential depicted in Figure 4 of Ref.~\cite{Mat23} 
by fitting it to a sum of three Gaussian functions, following the 
same methodology applied in Refs.~\cite{Gon18,Lyu21}.

As a first exploratory step, we consider only the contribution of 
the attractive $^1S_0$ $\Omega^i\Omega^i$ interaction. Within this 
one-channel approximation, no three-body bound state with 
$J^P = 3/2^+$ is found in any of the investigated flavor sectors. 
This notable outcome 
is a direct manifestation of the Pauli principle at the baryonic 
level. Specifically, the recoupling coefficient between the two spin-zero 
Faddeev amplitudes is negative (see Table~\ref{tab1}), which effectively 
converts the attractive two-body interaction into a repulsive one in 
the three-body system~\cite{Gar87}. 
Interestingly, enhancing the strength of the attraction in the $^1S_0$ 
channel further amplifies this effective repulsion in the three-body 
configuration. This behavior is a generic feature of systems composed 
of three identical fermions and is, in particular, known to apply 
to the three-neutron system. 
However, unlike neutrons, which are spin $1/2$ fermions, 
the $\Omega^i$ baryons 
are spin-$3/2$ particles. As a result, in addition to spin-zero two-body 
amplitudes, spin-two contributions are also allowed 
in the $L=0$ three-body state, as discussed above.
Crucially, the recoupling coefficient associated with the spin-two 
amplitudes is positive and thus unaffected by the 
antisymmetry constraints that effectively makes repulsive 
the spin-zero channel. This makes the inclusion of the spin-two 
component essential for a realistic description 
of the three-$\Omega^i$ system.

A comprehensive calculation of the $\Omega^i\Omega^i\Omega^i$ $S$-wave 
state with quantum numbers $J^P=3/2^+$ must account for contributions
from both relevant two-body partial 
waves: $^1S_0$ and $^5S_2$. A naive assumption of spin-independence in 
the $\Omega^i\Omega^i$ interaction~\cite{Gar24}, motivated, for instance,
by a Fermi--Breit-like potential at the baryonic level, can lead
to the prediction of deeply 
bound three-body states. In particular, when applying
the $^1S_0$ lattice QCD interaction from 
Ref.~\cite{Mat23} in the $i=b$ sector, such an approach yields binding
energies on the order of several hundred MeV. However, 
it is crucial to recognize that the $\Omega^i$ baryons are
identical particles, 
each composed of three identical quarks. As a result, the overall 
two-baryon wave function must remain antisymmetric under the exchange 
of quarks between baryons, particularly within their
spatial overlap region. This fundamental symmetry constraint 
limits the validity of spin-independent models and must be 
properly incorporated to ensure a physically meaningful 
description of the system.

The consequences of having identical quark constituents in the baryons
manifest themselves in the normalization of the 
two-baryon wave function~\cite{Val05}, 
which plays a crucial role in the calculation of observables,
most notably, in the effective interaction potential. The 
effect is encoded in the normalization kernel, which can be 
formally expressed as~\cite{Gar25}
\begin{eqnarray}
{\cal N}_{\Omega^i\Omega^i}^{LS} (R)  
\stackrel[R\to 0]{}{\hbox to 20pt{\rightarrowfill}}
&& 4 \pi \left[ {1 - {\frac{3 R^2}{4 \alpha^2}}} \right]
\frac{1}{1 \cdot 3 \cdots (2L+1)} \, 
\left[ \frac{R^2}{4 \alpha^2} \right]^L \label{norm2}\\
&& \times  \left\lbrace \left[ 3^L - 3 \, C(S) \right]
+  \frac{\left[ 3^{L+2} - 3 \, C(S) \right]}{2(2L+3)} \left[ \frac{R^2}{4 \alpha^2}
\right]^2  
+ \cdots \right\rbrace \, . \nonumber
\end{eqnarray}
%

$C(S)$ is a flavor-independent spin coefficient given by
\begin{equation}
C(S) =  \left< \Omega^i (123) \, , \Omega^i (456) ; S \right| P_{36}^{S}
\left| \Omega^i (123) \, ,\Omega^i (456); S \right> \, .
\end{equation}

The spin coefficients for the various $\Omega^i\Omega^i$ 
two-body states are presented in Table~\ref{tab2}.
\vspace{-6pt}
\begin{table}[H]
\caption{$C(S)$ {spin} coefficients for the $\Omega^i\Omega^i$ states.}
\begin{tabularx}{\textwidth}{CCCCC}
\toprule
 \boldmath{$S$ }   & \textbf{0}  & \textbf{ 1} &  \textbf{2} &  \textbf{3}  \\ \midrule
$C(S)$  &   $-\frac{1}{3}$ &   $-\frac{1}{9 }$ & $\frac{1}{3}$ &   $1$ \\
\bottomrule
\end{tabularx}
\label{tab2}
\end{table}

As seen from Equation~\eqref{norm2}, in $S$-wave configurations, 
the closer the coefficient $C(S)$ is to $1/3$, the more strongly 
the normalization of the two-baryon wave function 
is suppressed at short distances. This suppression reflects an 
effective Pauli repulsion, which originates from the 
presence of a quasi-forbidden state~\cite{Oka87}.

To derive the $^5S_2$ $\Omega^i\Omega^i$ interaction from the 
underlying quark dynamics, we employed the Born--Oppenheimer 
approximation~\cite{Lib77,Oka84,Val05}. Within this framework, the 
effective baryon--baryon potential is defined as
\begin{equation}
V_{B^\alpha B^\beta (L S ) \rightarrow B^\alpha B^\beta (L^{\prime} S^{\prime})} (R) =
\xi_{L \,S}^{L^{\prime}\, S^{\prime}} (R) \, - \, \xi_{L \,S}^{L^{\prime}\, S^{\prime}} (\infty) \, ,  \label{Poten1}
\end{equation}
\noindent where
\begin{equation}
\xi_{L \, S}^{L^{\prime}\, S^{\prime}} (R) \, = \, {\frac{{\left
\langle \Psi_{B^\alpha B^\beta}^{L^{\prime}\, S^{\prime}} ({\vec R}) \mid
\sum_{i<j=1}^{6} V_{q_iq_j}({\vec r}_{ij}) \mid \Psi_{B^\alpha B^\beta}^{L \, S} ({\vec R%
}) \right \rangle} }
{{\cal N}_{B^\alpha B^\beta}^{L^\prime S^\prime} (R)  \,\,
{\cal N}_{B^\alpha B^\beta}^{L S} (R)}} \, .
\label{Poten2}
\end{equation}

By inserting the spin coefficients from Table~\ref{tab2} into the
normalization expression
of the two-baryon wave function in Equation~\eqref{norm2}, we find that the 
$^5S_2$ $\Omega^i\Omega^i$ partial wave satisfies the condition 
$C(S) = \frac{1}{3}$. This value leads to a significant 
suppression of the wave function normalization at short distances,
manifesting as a strong 
Pauli repulsion. Crucially, this repulsive behavior arises from the
quark-level antisymmetrization constraints and cannot be captured 
by baryonic models alone. Consequently, a prominent short-range
repulsive core is predicted in the $^5S_2$ $\Omega^i\Omega^i$ 
interaction across all flavor sectors, purely as a consequence
of the quark substructure regardless of the specific details
of the interaction dynamics~\mbox{\cite{Oka81,Oky81}}.
In contrast, the $^1S_0$ $\Omega^i\Omega^i$ partial wave does not 
fulfill the Pauli blocking condition, i.e., $C(S) = 1/3$ (see 
Table~\ref{tab2}). As a result, its wave function normalization 
remains essentially constant at short distances,
indicating that no additional Pauli-induced repulsion arises beyond
what is already present in the interaction 
dynamics. This conclusion aligns well with existing lattice QCD 
results~\cite{Gon18,Lyu21,Mat23}, which also show that the $^1S_0$ 
channel exhibits genuine attractive behavior not masked by 
antisymmetrization effects.

\begin{figure}[H]
\includegraphics[width=.5\columnwidth]{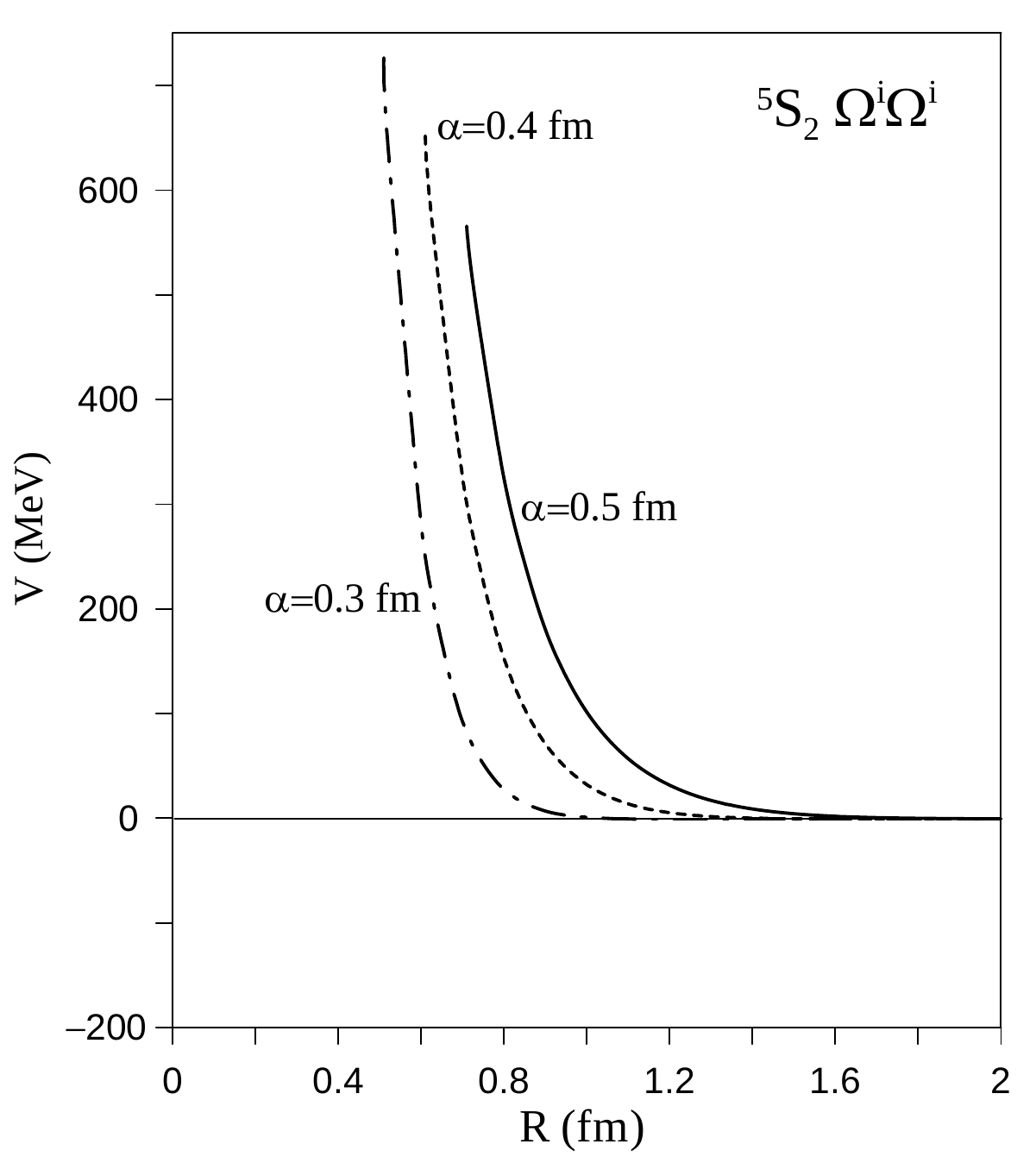}
\caption{$^5S_2$ $\Omega^i\Omega^i$ {interaction} 
{for} 
 different values of the Gaussian parameter of the quark wave function, $\alpha$.}
\label{fig22}
\end{figure}

The repulsive character of the $^5S_2$ $\Omega^i\Omega^i$ interaction is 
illustrated in Figure~\ref{fig22}, based on 
the parametrization and potential model of 
Ref.~\cite{Gar19}. The dominant contribution to this repulsion
originates from the Coulomb-like 
term of the one-gluon exchange potential, which is inherently 
flavor-independent. As a result, the spatial range 
of the repulsive core is determined by 
the Gaussian parameter $\alpha$ of the quark wave function, which 
characterizes the spatial extension of the $\Omega^i$ 
baryons~\cite{Gal14}. 
This behavior is, in fact, expected: constructing a total spin-2
state from six identical spin-1/2 quarks
of the same flavor (three per $\Omega^i$ baryon) while 
satisfying the color-singlet constraint 
is not possible
within the ground-state configuration. 
At least two quarks would be forced to occupy the 
same quantum state, in direct violation of the Pauli principle.
Therefore, this strong short-range repulsion is a model-independent
result, arising as a fundamental consequence of the antisymmetry
requirements at the quark level.

This repulsive effect, as previously reported in earlier 
works~\cite{Oky81}, is a generic feature of baryonic systems
subject to Pauli blocking. It manifest as a short-range
repulsive core at distances of approximately 
0.75--0.78~fm in light-quark systems, corresponding to 
a Gaussian parameter $\alpha \simeq 0.5$~fm. 
Crucially, such repulsion does not arise unless the 
antisymmetry of the baryon constituents is explicitly imposed. 
This is not the case in phenomenological approaches that disregard
the underlying quark substructure, such as those extrapolating 
from unrelated two-baryon systems or from the same system 
in quantum-number sectors unaffected by Pauli constraints. 
Ref.~\cite{Gar24} provides a clear example of this issue:
assuming spin independence 
in the $\Omega^i\Omega^i$ interaction while using 
the lattice QCD results of Ref.~\cite{Mat23}
yields artificially deeply bound 
states, with binding energies on the order of 
several hundred MeV in the $i = b$ sector.
Even if a strong repulsive core is introduced in the 
$^5S_2$ $\Omega^i\Omega^i$ channel
but with the same effective range as the $^1S_0$ channel 
(i.e., neglecting the full effect of Pauli blocking), 
the resulting three-body binding 
energy varies only mildly. This still leads to bound states with 
energies in the range of 250--350~MeV for systems with $i = b$.

Additional support for these conclusions comes from preliminary lattice 
QCD studies of $S$-wave scattering between strangeness $-3$ 
baryons~\cite{Buc12}. Conducted at a pion mass of $m_\pi \sim 390$~MeV 
across two lattice volumes, these studies indicate that the 
$^1S_0$ $\Omega\Omega$ interaction is 
at most weakly attractive, yielding an extrapolated scattering length 
of $a^{\Omega\Omega}_{S=0} = 0.16 \pm 0.22$~fm. In contrast, states 
in the $S=2$ channel appear at significantly higher energies, 
suggesting strong repulsion. Notably, it was proposed that none of the 
observed states correspond to the $S=2$ ground state, highlighting 
the need for future simulations at lower pion masses.

To assess the combined effect of both spin channels, 
we solved the full three-body $\Omega^i\Omega^i\Omega^i$ 
problem, incorporating both the $^1S_0$ 
and $^5S_2$ two-body interactions. The $^1S_0$ interactions were taken 
from lattice QCD results~\cite{Gon18,Lyu21,Mat23}, while the $^5S_2$ 
interaction corresponds to the strongly repulsive potential
shown in Figure~\ref{fig22}. The repulsive character of 
the single-channel calculation, driven by Pauli effects at the 
baryon level, is significantly enhanced by the quark-level
antisymmetry that dominates the $^5S_2$ channel. 
As a result, despite the attractive nature of the $^1S_0$
interaction in the two-body sector, we find no evidence of a 
$J^P = 3/2^+$ bound state in the $\Omega^i\Omega^i\Omega^i$ 
system for any flavor sector.

\section{Decay Width of a Resonance Between Two Thresholds}
\label{secIV}

A recurring feature among many of the proposed multiquark candidates 
discussed in the preceding sections is their theoretical 
interpretation as bound states of higher-energy 
configurations situated near the observed mass yet
experimentally detected in lower-energy decay channels~\cite{Liu19}.
This leads to a nontrivial relationship between the 
decay width and the available phase space, challenging the 
conventional explanation that larger phase spaces
generally result in broader widths. A notable example is provided 
by the $P_c^+$ pentaquark states discussed in Section~\ref{secII}. 
Table~\ref{Tab_15} presents the phase space available for 
the $J/\psi \, p$ decay, along with the corresponding widths. 
Interestingly, the state with the largest phase space exhibits the 
narrowest decay width, whereas other states with nearly identical phase 
space show widths that differ by orders of magnitude. 
This counterintuitive 
behavior is both intriguing and concerning, as it highlights
the limitations of a purely kinematic interpretation 
and suggests the presence of more intricate dynamical mechanisms
at play. Similar patterns have been observed in the meson sector, 
reinforcing that this phenomenon is not exclusive to baryonic 
resonances. A comprehensive discussion, along with additional
examples, is available in Ref.~\cite{Liu19} 
and the references therein.
\vspace{-3pt}
\begin{table}[H]
\caption{{Summary} of the $P_c^+$ properties decaying to $J/\psi \, p$~\cite{Aai15,Aai19}.}
\begin{tabularx}{\textwidth}{CCCC}
\toprule
 \textbf{State}         & \boldmath{$M$} \textbf{(MeV)}                       & \boldmath{$\Gamma$} \textbf{(MeV) }                 & \textbf{Phase Space (MeV)}    \\ \midrule
 $P_c(4380)^+$ & $4380 \pm 8 \pm 29$             & $205 \pm 18 \pm 86$             &  345 $\pm$ 30        \\
$P_c(4312)^+$ & $4311.9 \pm 0.7 ^{+6.8}_{-0.6}$ & $9.8 \pm 2.7 ^{+3.7}_{-4.5}$    &  276.7 $\pm$ 6.8     \\  $P_c(4440)^+$ & $4440.3 \pm 1.3 ^{+4.1}_{-4.7}$ & $20.6 \pm 4.9 ^{+8.7}_{-10.1}$  &  405.1 $\pm$ 4.9     \\ 
 $P_c(4457)^+$ & $4457.3 \pm 0.6 ^{+4.1}_{-1.7}$ & $6.4 \pm 2.0 ^{+5.7}_{-1.9}$    &  422.1 $\pm$ 4.1     \\ 
\bottomrule
\end{tabularx}
\label{Tab_15} 
\end{table}
\vspace{-3pt}

If the quark dynamics within a $Q\bar Q q \bar q$ 
multiquark system is not constrained
by the presence of correlated substructures, then two distinct 
color-singlet configurations can contribute to the total wave function,

\begin{equation}
\Psi_{4q} = \alpha_1 \left| [Q\bar Q][q\bar q] \right\rangle + \alpha_2 \left|[Q\bar q][q\bar Q]\right\rangle \, ,
\label{Eq_1}
\end{equation}
where, in general, $\alpha_1$ and $\alpha_2$, determined by the 
underlying quark dynamics, are both nonzero. A similar decomposition
holds for $Q\bar Q q q q$ multiquark baryons,
\begin{equation}
\Psi_{5q} = \beta_1 \left| [Q\bar Q][qqq] \right\rangle + \beta_2 \left|[q\bar Q][Qqq]\right\rangle \, .
\label{Eq_2}
\end{equation}

The properties of such states should be addressed by means of a 
coupled-channel calculation, as detailed in Section~\ref{subsecIII.II}. 
Consequently, it becomes relevant to analyze the masses of the various 
thresholds and identify the dominant color-singlet components across 
different flavor sectors. This threshold structure not only constrains 
the dynamics of the system but also governs the coupling strength 
between channels, thereby playing a central role in determining the 
formation, binding mechanism, and decay patterns of the multiquark states.

\begin{figure}[H]
\includegraphics[width=0.7\columnwidth]{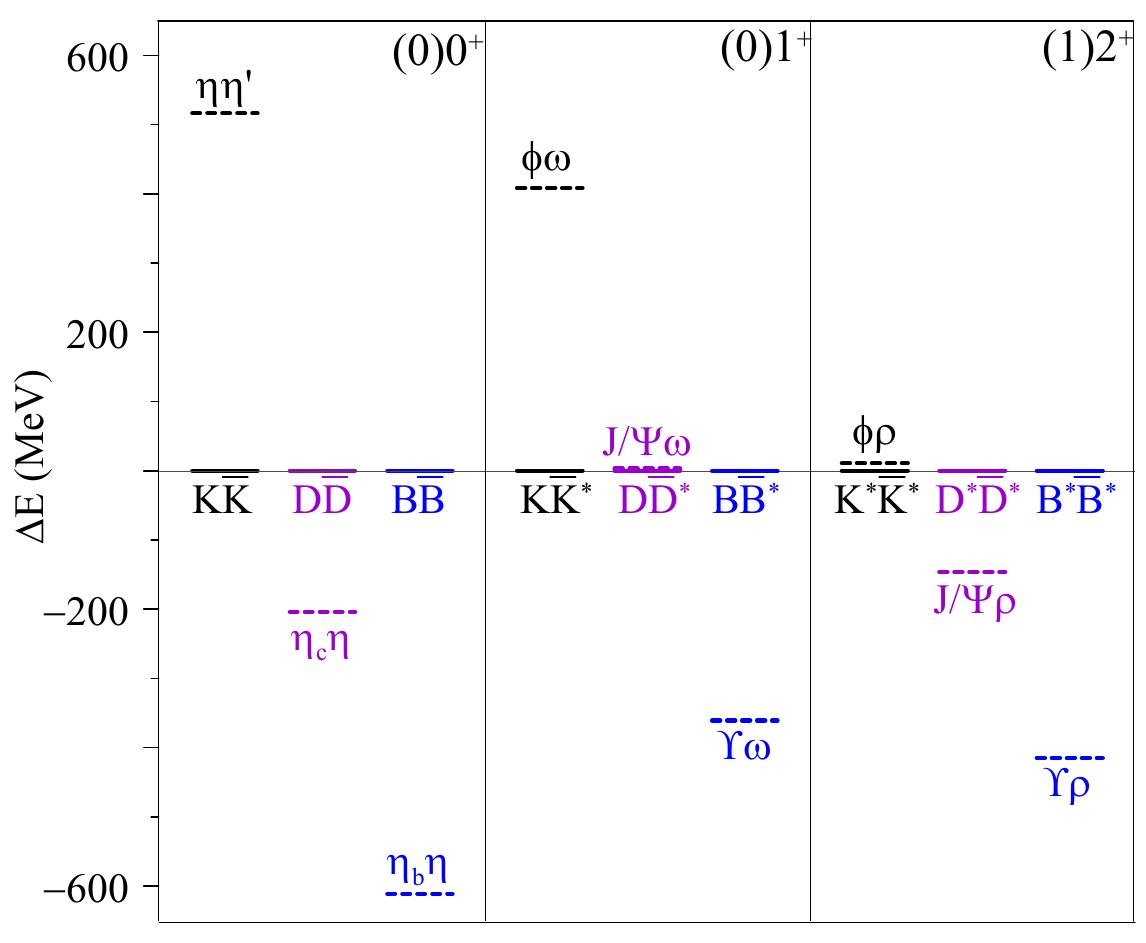}
\caption{{Experimental} 
 masses~\cite{Nav24} of the different color singlets
that make up $Q\bar Q q\bar q$ multiquarks with $Q=s$, $c$, or $b$ 
for different sets of quantum numbers $(I)J^{P}$. 
The reference energy has been set to the corresponding 
$K\bar K$, $D\bar D$, and $B\bar B$ mass for the hidden
strange, charm, and bottom \mbox{sectors,~respectively.}}
\label{Fig_4}
\end{figure}

\begin{figure}[H]
\includegraphics[width=0.7\columnwidth]{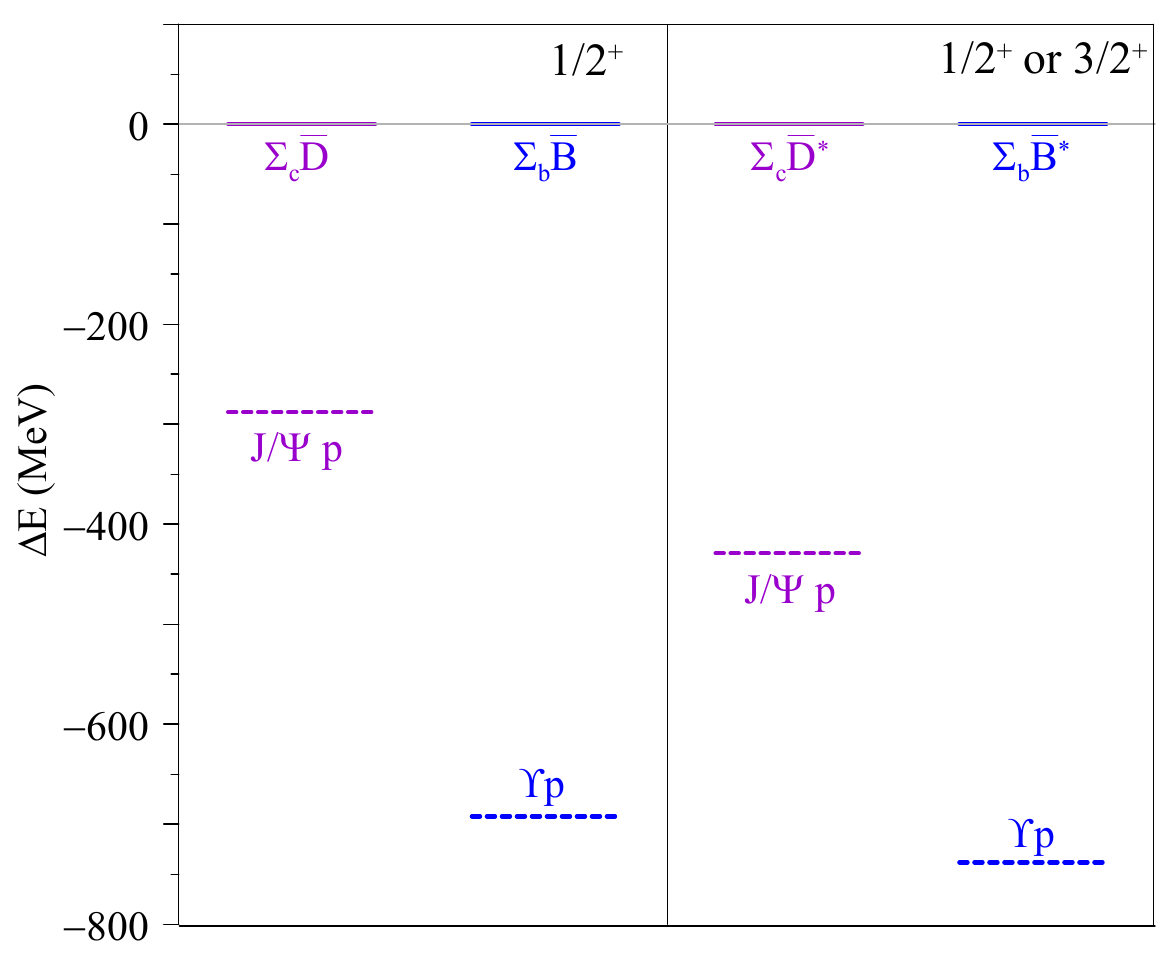}
\caption{{Experimental} 
 masses~\cite{Nav24} of the different color singlets
that make up selected $Q\bar Q qqq$ multiquarks with $Q=c$ or $b$ 
for different sets of quantum numbers $J^{P}$. The 
reference energy has been set to the corresponding 
$\Sigma_c\bar D$ and $\Sigma_b \bar B$ mass for the hidden
charm and bottom sectors,~respectively.}
\label{Fig_5}
\end{figure}

Multiquark systems with a flavor content of $Q\bar Q q\bar q$ 
can form two types of color-singlet configurations: 
$[Q \bar q][q \bar Q]$ and
$[Q \bar Q][q \bar q]$. For the $Q=c$ case, the masses of these two
configurations are relatively close. In contrast,
for $Q=b$, the $[Q \bar Q][q \bar q]$ state lies significantly
lower in mass, as shown in Figure~\ref{Fig_4}. In the strange sector,
the situation is reversed, where 
the $[Q \bar q][q \bar Q]$ component becomes the lowest in energy, 
potentially leading to the formation of stable meson--antimeson 
molecular {states} 
~\cite{Wei90}.
A similar pattern arises in multiquark systems 
of the form $Q\bar Q qqq$. In the charm
sector, for spin $J=1/2$, the mass difference 
between the two possible color-singlet configurations is
$\Delta M = M_{[q\bar Q][Qqq]} - 
M_{[Q\bar Q][qqq]} = M_{\bar D \Sigma_c} - M_{J/\psi \, p} =$ 288~MeV. 
However, in the bottom sector, this difference increases significantly
to $\Delta M = M_{[q\bar Q][Qqq]} - 
M_{[Q\bar Q][qqq]} = M_{\bar B \Sigma_b} - 
M_{\Upsilon p} =$ 692~MeV~\cite{Nav24}.
This behavior is illustrated in Figure~\ref{Fig_5} for 
multiquark baryons with quantum numbers $J^P=1/2^+$ and 
$J^P=3/2^+$ in both the charm and bottom sectors.
Several of the proposed multiquark states
correspond to resonances that lie between the masses of the
two color-singlet configurations.

\begin{figure}[H]
\includegraphics[width=0.75\columnwidth]{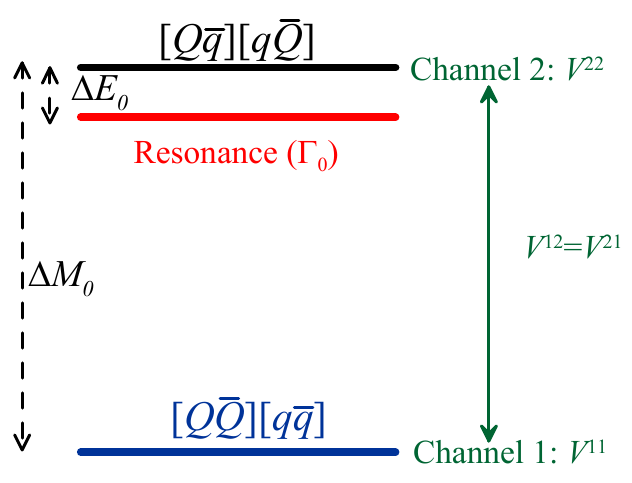}
\caption{Diagrammatic representation of the modeled 
experimental scenario for $B=0$.
An analogous diagram applies to the $B=1$ sector,
with the substitution of $[Q\bar q][q\bar Q]$ by
$[q\bar Q][Qqq]$ and $[Q\bar Q][q\bar q]$ by $[Q\bar Q][qqq]$.
See text for further details.}
\label{Fig_6}
\end{figure}

The scenario described above can be modeled using
a coupled-channel framework involving two distinct
two-body color-singlet configurations, with a resonance 
emerging between them, as illustrated in Figure~\ref{Fig_6}.
Channel 1, which is lower in mass, consists of two color 
singlets: $[Q\bar Q][q\bar q]$ for systems with baryon number $B=0$
and $[Q\bar Q][qqq]$ for $B=1$. We denote
the mass of the $Q\bar Q$ state as $m_1$ and 
that of the $q \bar q$ (or $qqq$) state as $m_2$.
Channel 2, which is higher in mass, also comprises 
two color singlets: $[Q\bar q][q\bar Q]$ for $B=0$
and $[q\bar Q][Qqq]$ for $B=1$.
The mass of the $Q\bar q$ state is denoted by $m_3$ 
and that of the $q \bar Q$ (or $Qqq$) state by $m_4$.

\vspace{-12pt}

Specifically, we model the system as a coupled-channel 
problem governed by the 
non-relativistic Lippmann--Schwinger equation,
where the two-body potentials
consist of a combination of attractive and repulsive Yukawa terms, i.e.,

\begin{equation}
V^{ij}(r)=-A^{ij} \frac{e^{-\mu^{ij}_Ar}}{r} + B^{ij} \frac{e^{-\mu^{ij}_Br}}{r}.
\label{Eq_3} 
\end{equation}

We consider scenarios in which a resonance occurs at an 
energy $E=E_R$, defined by the condition that the scattering 
phase shift satisfies $\delta(E_R)=90^\circ$,
corresponding to a Breit--Wigner resonance.
This resonances lies between the mass thresholds of channels 1 and 2,
i.e., $0 < E_R < \Delta M$. The physical mass of the 
resonance is then given by $W_R=E_R + m_1 +m_2$.
The width of the resonance
is calculated using the Breit--Wigner formula, as
described in Refs.~\cite{Bre36,Cec08,Cec13},
\begin{equation}
\Gamma (E) =\lim\limits_{E \to E_R}\, \frac{2(E_R-E)}{\text{cotg}[\delta(E)]} \, .
\label{Eq_4} 
\end{equation}

We begin by requiring that in a single-channel calculation, 
the upper channel \mbox{(channel~2)} supports
a bound state with energy close to zero. In contrast, when both
channels are coupled, the system develops a bound state just 
below the threshold of the lower channel (channel 1). 
By increasing, for instance, the strength of the repulsive 
component in the lower channel, this bound state 
can be shifted upward, eventually becoming a 
resonance embedded in the continuum.
This framework allows for the 
study of how the resonance width evolves as its mass 
transitions from the vicinity of the lower channel 
to that of the upper channel. The initial configuration 
is illustrated in Figure~\ref{Fig_6}, representing a resonance 
near the upper-channel threshold, corresponding to the parameters 
listed in Table~\ref{Tab_16}. Numerically, $\Delta E_0 =$ 3.64 MeV,
\mbox{$\Gamma_0 =$ 9.6 MeV}, and $\Delta M_0 =$ 25.6~MeV.
\begin{table}[H]
\caption{{Parameters} of the interaction as defined 
in Equation~\protect\eqref{Eq_3}. The coefficients $A^{ij}$ and $B^{ij}$ 
are given in MeV$\cdot$fm, while the inverse ranges $\mu^{ij}_A$ 
and $\mu^{ij}_B$ are expressed in ${\rm fm}^{-1}$.} 
\begin{tabularx}{\textwidth}{CCCCC}
\toprule
\boldmath{$V^{ij}$  }       & \boldmath{ $A^{ij}$ }   & \boldmath{$\mu^{ij}_A$ }& \boldmath{$B^{ij}$}   & \boldmath{$\mu^{ij}_B$}  \\ \midrule
$V^{11}$          &  $100$       & $2.68$       & $3000$     & $5.81$        \\ 
$V^{22}$          &  $680$       & $4.56$       & $642$      & $6.73$        \\ 
$V^{12}=V^{21}$   &  $200$       & $1.77$       & $195$      & $3.33$        \\ 
\bottomrule
\end{tabularx}
\label{Tab_16} 
\end{table}
\vspace{-3pt}

\begin{figure}[H]
{\includegraphics[width=0.75\columnwidth]{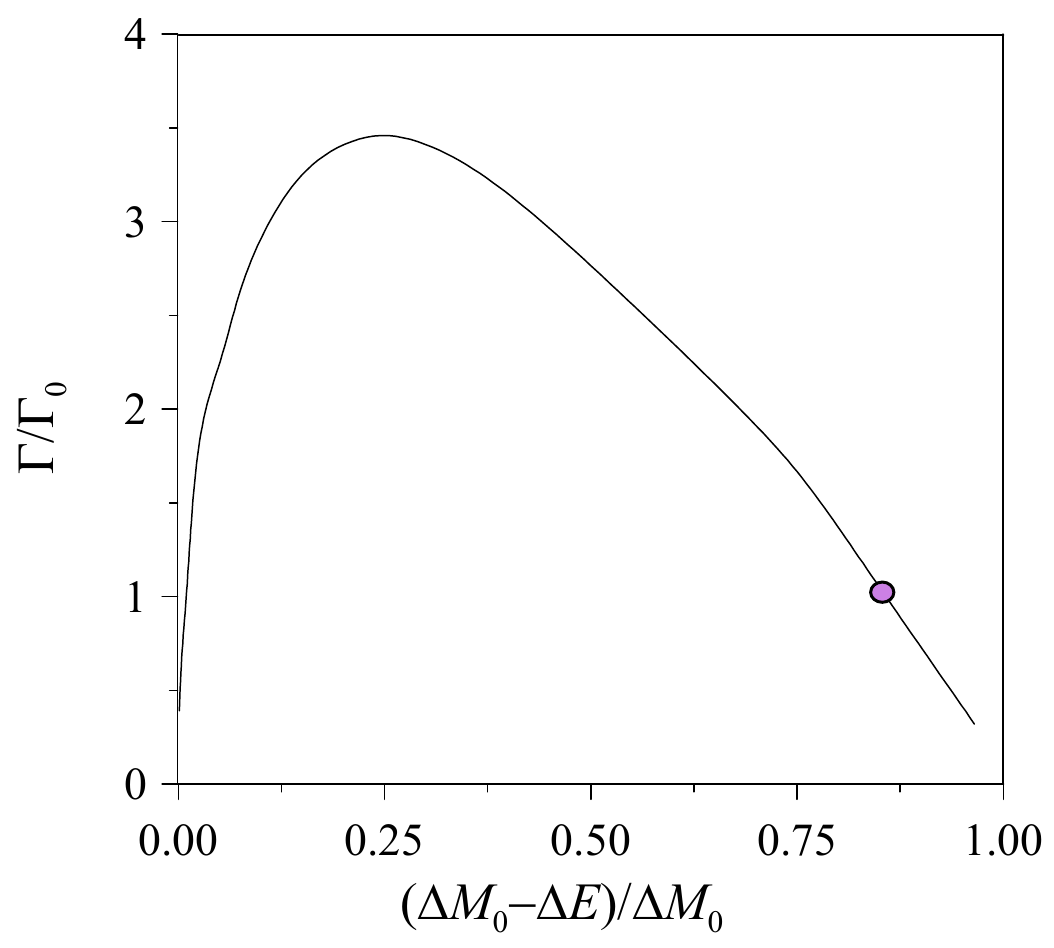}}
\caption{\textls[-15]{Variation in the resonance decay width 
$\Gamma$ (in units of $\Gamma_0$)
as a function of its relative position between the 
formation and detection channels, expressed as 
$\Delta M_0 - \Delta E$ (in units of $\Delta M_0$). 
The purple circle marks the starting point corresponding to
the configuration shown in Figure~\ref{Fig_6}. See text for details.}}
\label{Fig_8}
\end{figure}

Let us examine how the decay width of the resonance evolves
as its position shifts between the lower, $\Delta E \to \Delta M_0$,
an the upper, $\Delta E \to 0$, channels. 
In other words, both the phase space available for decay 
and the binding energy relative to the formation channel 
change simultaneously. Since
the resonance is generated by the attraction in the upper channel, 
we vary the repulsion in the detection (lower) channel to 
adjust its position. 
The results, shown in Figure~\ref{Fig_8}, reveal how the 
resonance width initially increases rapidly as it approaches 
the lower channel despite the decreasing 
phase space. However, roughly two-thirds of the way toward 
the lower channel, the width begins to decrease.
This behavior highlights an important point:
for resonances significantly separated from their detection 
channel, it is the mass difference relative to the formation channel
that predominantly governs the resonance width. Specifically,
a larger binding energy corresponds to a 
larger width, even though the available phase space diminishes.
As the resonance nears the upper channel, it becomes narrow and
appears to largely ignore 
the lower channel. This is because the wave function of the 
nearly zero-energy bound in channel 2 has minimal overlap with 
the configuration of channel 1. Consequently, 
in this region, the dynamics are dominated by the attraction 
within the upper channel, with the second channel serving 
primarily as a detection mechanism.

The results presented in Figure~\ref{Fig_8} align well with
recent observations by the LHCb Collaboration. 
In Table~\ref{Tab_17}, we summarize the most commonly proposed
molecular interpretations of the $P_c^+$ pentaquarks.
Focusing first on the latest LHCb results, listed in the last 
three rows of Table~\ref{Tab_17}, we note that these resonances 
lie far from their detection channel, $J/\psi \, p$,
with the corresponding phase space values given in Table~\ref{Tab_15}.
As demonstrated in Figure~\ref{Fig_8}, when a resonance is distant 
from its detection channel, the key factor determining its decay 
width is the mass difference relative to the formation channel.
The binding energies of the recent $P_c^+$ states increase as follows 
(see Table~\ref{Tab_17}): 2.5~MeV, 5.8~MeV, and 19.5~MeV.
Correspondingly, their observed decay widths 
(from Table~\ref{Tab_15}) are 6.4~MeV, 9.8~MeV, 
and~20.6 MeV, respectively, showing excellent 
agreement with the trends predicted by our model.

\vspace{-6pt}
\begin{table}[H]
\caption{{Suggested} molecular structures for the $P_c^+$ 
pentaquarks~\cite{Aai15,Aai19,Bur15,Kar21}.} 
\begin{tabularx}{\textwidth}{CcCC}
\toprule
\textbf{ State}         & \textbf{Suggested Molecule  }      & \boldmath{$M_{\rm Molecule}$} \textbf{(MeV)}& \boldmath{$\Delta E$} \textbf{(MeV)  }   \\ \midrule
 $P_c(4380)^+$ & ${\Sigma_c^*}^+ \bar D^0$ & $4382.3 \pm 2.3$    & 2.3  $\pm$ 30.1   \\
 $P_c(4312)^+$ & $\Sigma_c^+ \bar D^0$     & $4317.7 \pm 0.4$    & 5.8  $\pm$  6.8   \\ 
 $P_c(4440)^+$ & $\Sigma_c^+ \bar {D^*}^0$ & $4459.8 \pm 0.4$    & 19.5 $\pm$ 4.9    \\
 $P_c(4457)^+$ & $\Sigma_c^+ \bar {D^*}^0$ & $4459.8 \pm 0.4$    & 2.5  $\pm$ 4.1    \\ 
 \bottomrule
\end{tabularx} 
\label{Tab_17} 
\end{table}

\section{Summary and Outlook}
\label{secV}
We present a summary of our recent findings on the structure and 
properties of exotic hadrons, focusing on two complementary research 
directions.

The first line of investigation addresses hidden-flavor 
pentaquark states, incorporating dynamical correlations 
arising from the Coulomb-like behavior 
of the short-range color interaction. These correlations 
effectively ``freeze'' the color wave function, 
allowing the reduction of 
the original five-body problem to a simplified three-body system. 
This reduced system is solved exactly using Faddeev equations.
Our results provide a coherent description of the observed spectrum in 
the hidden-charm sector and offer predictive insights for the 
hidden-bottom sector, where experimental data are still unavailable. 
It is important to note that the color correlations employed here 
do not universally lead to bound multiquark states for all quark 
configurations. In particular, for pentaquarks containing anticharm
or anti bottom quarks, the color wave function remains unconstrained 
within this framework. Finally,
our model yields a dynamical mechanism for the possible formation 
of quarkonium bound to nuclear matter, an outcome not previously 
derived from quark--gluon dynamics within a restricted Hilbert 
space.

The second part of our study explores recent developments from lattice 
QCD and phenomenological analyses that suggest the potential 
stability of hadronic molecules containing multiple heavy quarks. 
These results indicate that few-body systems 
involving heavy flavors may inherently favor bound-state
formation. As illustrative examples, we consider 
extreme configurations such as multiquark systems incorporating 
the doubly bottom tetraquark $T_{bb}$ or the $\Omega\Omega$ 
dibaryon states predicted by lattice QCD.
In three-meson systems, we observe a general trend in which
the binding energy diminishes with an increasing number of 
hadronic constituents. This decrease arises from the combined
effects of the
Pauli principle and the emergence of additional kinematic 
thresholds. Furthermore, the binding is weakened as 
the heavy-quark mass is reduced, due to the increasing mass
differences among internal two-body subsystems, which
act to destabilize potential bound configurations.
Turning to three-baryon systems, we find that the Pauli principle 
again plays a critical role, particularly in 
suppressing the formation of bound states composed of three 
$\Omega^i$ baryons. Two dominant effects 
are identified. First, for spin-zero two-body subsystems, 
a negative spin recoupling coefficient transforms attractive $^1S_0$ 
interactions into effectively repulsive ones when extended to 
the full three-body system. Second, while spin-two configurations
are not constrained by the Pauli principle 
at the baryon level, they still encounter 
strong short-range repulsion at the 
quark level due to antisymmetrization. Together, these 
effects significantly hinder the formation of 
$S$-wave three-$\Omega$ bound 
states regardless of the specific flavor content.

Finally, we examine the decay width of multiquark resonances 
composed of multiple color-singlet substructures.
We have illustrated how narrow resonances can exist 
even when a large phase space for decay is available. In particular,
for resonances located far from the 
experimental detection threshold, the decay width is primarily 
determined by the mass difference between the resonance 
and its formation channel 
rather than by the available phase space in the decay channel.
As a result, more deeply bound states can exhibit 
broader widths despite having limited phase space for decay.
We also explore the potential existence of multiquark analogs 
across different flavor sectors. Our analysis reveals that 
structural similarities do not necessarily translate 
into similar physical properties.
This highlights the importance of caution when 
extrapolating multiquark predictions across flavor sectors, 
especially in cases where the underlying dynamics are 
approximately flavor-independent and 
dominated by coupled-channel effects.

Exotic hadron spectroscopy remains a powerful testing 
ground for deepening our understanding of QCD in the 
non-perturbative regime. Despite the inherent complexity
of the subject, this review has outlined several 
mechanisms that quantitatively account for a variety of compelling 
experimental observations. While significant challenges persist,
continued progress hinges on the development of theoretical 
frameworks capable of uncovering the dynamics that govern
low-energy hadron dynamics. We 
hope that this work provides a meaningful 
contribution to that ongoing endeavor.
\vspace{6pt} 





\authorcontributions{Conceptualization, H.G. and A.V.; methodology, H.G. and A.V.;
formal analysis, H.G.; investigation, H.G. and A.V.; writing-original
draft preparation, A.V.; writing-review and editing, H.G. and A.V.; 
funding acquisition, H.G. and A.V.  All authors have read and agreed to
the published version of the manuscript.}

\funding{{This work has been partially funded by COFAA-IPN (M\'exico),
by the Spanish Ministerio de Ciencia e
Innovaci\'on (MICINN) and the European Regional Development Fund (ERDF),
under contract RED2024-153978-E; by
Spanish MICIU/AEI/10.13039/501100011033 
grant No. PID2022-141910NB-I00; and by
Junta de Castilla y Le\'on under contract SA091P24.}}

\dataavailability{The original contributions presented in this study are included in the article. Further inquiries can be directed to the corresponding author(s).}

\conflictsofinterest{{The authors declare no conflict of interest.}}
\begin{adjustwidth}{-\extralength}{0cm}

\reftitle{References}

\PublishersNote{}
\end{adjustwidth}

\begin{thebibliography}{999}
\bibitem{Gel64} Gell-Mann, M.
A Schematic Model of Baryons and Mesons.
\emph{Phys. Lett.} \textbf{1964}, \emph{8}, 214--215. [\href{http://doi.org/10.1016/S0031-9163(64)92001-3}{CrossRef}]

\bibitem{Bar95} Barnes, T.; Close, F.E.; Swanson, E.S.
Hybrid and conventional mesons in the flux tube model: Numerical studies and their phenomenological implications.
\emph{Phys. Rev. D} \textbf{1995}, \emph{52}, 5242--5256. [\href{http://dx.doi.org/10.1103/PhysRevD.52.5242}{CrossRef}] [\href{http://www.ncbi.nlm.nih.gov/pubmed/10019746}{PubMed}]

\bibitem{Jaf77} Jaffe, R.L.
Multiquark hadrons. I. Phenomenology of $Q^2 \bar Q^2$ mesons.
\emph{Phys. Rev. D} \textbf{1977}, \emph{15}, 267--280. [\href{http://dx.doi.org/10.1103/PhysRevD.15.267}{CrossRef}]

\bibitem{Jae77} Jaffe, R.L.
Multi-Quark Hadrons. 2. Methods.
\emph{Phys. Rev. D} \textbf{1977}, \emph{15}, 281--289. [\href{http://dx.doi.org/10.1103/PhysRevD.15.281}{CrossRef}]

\bibitem{Vac05} Valcarce, A.; Garcilazo, H.; Vijande, J.
Constituent quark model study of light- and strange-baryon spectra.
\emph{Phys. Rev. C} \textbf{2005}, \emph{72},~025206. [\href{http://dx.doi.org/10.1103/PhysRevC.72.025206}{CrossRef}]

\bibitem{Mag05} Magas, V.K.; Oset, E.; Ramos, A.
Evidence for the Two-Pole Structure of the $\Lambda(1405)$ Resonance.
\emph{Phys. Rev. Lett.} \textbf{2005}, \emph{95}, 052301. [\href{http://dx.doi.org/10.1103/PhysRevLett.95.052301}{CrossRef}]

\bibitem{Cap86} Capstick, S.; Isgur, N.
Baryons in a relativized quark model with chromodynamics.
\emph{Phys. Rev. D} \textbf{1986}, \emph{34}, 2809--2835. [\href{http://dx.doi.org/10.1103/PhysRevD.34.2809}{CrossRef}]

\bibitem{Cle17} Clement, H.
On the History of Dibaryons and their Final Observation.
\emph{Prog. Part. Nucl. Phys.} \textbf{2017}, \emph{93}, 195. [\href{http://dx.doi.org/10.1016/j.ppnp.2016.12.004}{CrossRef}]

\bibitem{Tri04} Trilling, G.
Reviews of Particle Physics edited by Eidelman, S., Hayes, K.G., Olive, K.E., Aguilar-Benitez, M., Amsler, C., Asner, D., Babu, K.S., Barnett, R.M., Beringer, J., Burchat, P.R., et al. Review of particle physics.
\emph{Phys. Lett. B} \textbf{2004}, \emph{592}, 1--5.

\bibitem{Ade82} Ader, J.-P.; Richard, J.-M.; Taxil, P.
Do narrow heavy multiquark states exist?
\emph{Phys. Rev. D} \textbf{1982}, \emph{25}, 2370. [\href{http://dx.doi.org/10.1103/PhysRevD.25.2370}{CrossRef}]

\bibitem{Pic95} Pich, A.
Chiral perturbation theory.
\emph{Rep. Prog. Phys.} \textbf{1995}, \emph{58}, 563--610. [\href{http://dx.doi.org/10.1088/0034-4885/58/6/001}{CrossRef}]

\bibitem{Oll00} Oller, J.A.; Oset, E.; Ramos, A.
Chiral unitary approach to meson meson and meson-baryon interactions and nuclear applications.
\emph{Prog. Part. Nucl. Phys.} \textbf{2000}, \emph{45}, 157--242. [\href{http://dx.doi.org/10.1016/S0146-6410(00)00104-6}{CrossRef}]

\bibitem{Aai22} Aaij, R.; Abdelmotteleb, A.S.W.; Abell\'an Beteta, C.;
Abudinen Gallego, F.J.; Ackernley, T.; Adeva, B.;
Adinolfi, M.; Afsharnia, H.; Agapopoulou, C.; Aidala, C.A.; et al.
Observation of an exotic narrow doubly charmed tetraquark.
\emph{Nat. Phys.} \textbf{2022}, \emph{18}, 751--754. [\href{http://dx.doi.org/10.1038/s41567-022-01614-y}{CrossRef}]

\bibitem{Aaj22} Aaij, R.; Abdelmotteleb, A.S.W.; Abell\'an Beteta, C.;
Abudinen Gallego, F.J.; Ackernley, T.; Adeva, B.;
Adinolfi, M.; Afsharnia, H.; Agapopoulou, C.; Aidala, C.A.; et al.
Study of the doubly charmed tetraquark $T_{cc}^+$.
\emph{Nat. Commun.} \textbf{2022}, \emph{13}, 3351.

\bibitem{Jaf05} Jaffe, R.L.
Exotica.
\emph{Phys. Rep.} \textbf{2005}, \emph{409}, 1--45. [\href{http://dx.doi.org/10.1016/j.physrep.2004.11.005}{CrossRef}]

\bibitem{Che16} Chen, X.H.; Chen, W.; Liu, X.; Zhu, L.S.
The hidden-charm pentaquark and tetraquark states.
\emph{Phys. Rep.} \textbf{2016}, \emph{639}, 1--121. [\href{http://dx.doi.org/10.1016/j.physrep.2016.05.004}{CrossRef}]

\bibitem{Bri16} Brice\~no, R.A.; Cohen, T.D.; Coito, S.;
Dudek, J.J.; Eichten, E.; Fischer, C.S.;
Fritsch, M.; Gradl, W.; Jackura, A.; Kornicer, M.; et al.
Issues and Opportunities in Exotic Hadrons.
\emph{Chin. Phys. C} \textbf{2016}, \emph{40}, 042001. [\href{http://dx.doi.org/10.1088/1674-1137/40/4/042001}{CrossRef}]

\bibitem{Ric16} Richard, J.-M.
Exotic hadrons: Review and perspectives.
\emph{Few Body Syst.} \textbf{2016}, \emph{57}, 1185--1212. [\href{http://dx.doi.org/10.1007/s00601-016-1159-0}{CrossRef}]

\bibitem{Hos16} Hosaka, A.; Iijima, T.; Miyabayashi, K.; Sakai, Y.; Yasui, S.
Exotic hadrons with heavy flavors: X, Y, Z, and related states.
\emph{Prog. Theor. Exp. Phys.} \textbf{2016}, 062C01. [\href{http://dx.doi.org/10.1093/ptep/ptw045}{CrossRef}]

\bibitem{Che17} Chen, H.-X.; Chen, W.; Liu, X.; Liu, Y.-R.; Zhu, S.-L.
A review of the open charm and open bottom systems.
\emph{Rep. Prog. Phys.} \textbf{2017}, \emph{80}, 076201. [\href{http://dx.doi.org/10.1088/1361-6633/aa6420}{CrossRef}]

\bibitem{Leb17} Lebed, R.F.; Mitchell, R.E.; Swanson, E.S.
Heavy-Quark QCD Exotica.
\emph{Prog. Part. Nucl. Phys.} \textbf{2017}, \emph{93}, 143. [\href{http://dx.doi.org/10.1016/j.ppnp.2016.11.003}{CrossRef}]

\bibitem{Ali17} Ali, A.; Lange, J.S.; Stone, S.
Exotics: Heavy Pentaquarks and Tetraquarks.
\emph{Prog. Part. Nucl. Phys.} \textbf{2017}, \emph{97}, 123--198. [\href{http://dx.doi.org/10.1016/j.ppnp.2017.08.003}{CrossRef}]

\bibitem{Esp17} Esposito, A.; Pilloni, A.; Polosa, A.D.
Multiquark Resonances.
\emph{Phys. Rep.} \textbf{2017}, \emph{668}, 1--97. [\href{http://dx.doi.org/10.1016/j.physrep.2016.11.002}{CrossRef}]

\bibitem{Guo18} Guo, F.-K.; Hanhart, C.;  Mei\ss ner, U.-G.; Wang, Q.;
Zhao, Q.; Zou, B.-S.
Hadronic molecules.
\emph{Rev. Mod. Phys.} \textbf{2018}, \emph{90}, 015004. [\href{http://dx.doi.org/10.1103/RevModPhys.90.015004}{CrossRef}]

\bibitem{Ols18} Olsen, S.L.; Skwarnicki, T.; Zieminska, D.
Nonstandard heavy mesons and baryons: Experimental evidence.
\emph{Rev. Mod. Phys.} \textbf{2018}, \emph{90}, 015003. [\href{http://dx.doi.org/10.1103/RevModPhys.90.015003}{CrossRef}]

\bibitem{Kar18} Karliner, M.; Rosner, J.L.; Skwarnicki, T.
Multiquark States
\emph{Ann. Rev. Nucl. Part. Sci.} \textbf{2018}, \emph{68}, 17--44. [\href{http://dx.doi.org/10.1146/annurev-nucl-101917-020902}{CrossRef}]

\bibitem{Bra20} Brambilla, N.; Eidelman, S.; Hanhart, C.; Nefediev, A.;
Shen, C.-P.; Thomas, C.E.; Vairo, A.; Yuan, C.-Z.
The $XYZ$ states: experimental and theoretical status and perspectives.
\emph{Phys. Rep.} \textbf{2020}, \emph{873}, 1--154. [\href{http://dx.doi.org/10.1016/j.physrep.2020.05.001}{CrossRef}]

\bibitem{Yan20} Yang, G.; Ping, J.; Segovia, J.
Tetra- and penta-quark structures in the constituent quark model.
\emph{Symmetry} \textbf{2020}, \emph{12}, 1869. [\href{http://dx.doi.org/10.3390/sym12111869}{CrossRef}]

\bibitem{Hua23} \textls[-15]{Huang, H.; Deng, C.; Liu, X.; Tan, Y.; Ping, J.
Tetraquarks and Pentaquarks from Quark Model Perspective.
\emph{Symmetry} \textbf{2023}, \emph{15}, 1298. [\href{http://dx.doi.org/10.3390/sym15071298}{CrossRef}]}

\bibitem{Hug18} Hughes, C.; Eichten, E.; Davies, C.T.H.
Searching for beauty-fully bound tetraquarks using lattice nonrelativistic QCD.
\emph{Phys. Rev. D} \textbf{2018}, \emph{97}, 054505. [\href{http://dx.doi.org/10.1103/PhysRevD.97.054505}{CrossRef}]

\bibitem{Hud20} Hudspith, R.J.; Colquhoun, B.; Francis, A.; Lewis, R.; Maltman, K.
Lattice investigation of exotic tetraquark channels.
\emph{Phys. Rev. D} \textbf{2020}, \emph{102}, 114506. [\href{http://dx.doi.org/10.1103/PhysRevD.102.114506}{CrossRef}]

\bibitem{Col24} Colquhoun, B.; Francis, A.; Hudspith, R.J.; Lewis, R.;
Maltman, K.; Parrott, W.G.
Improved analysis of strong-interaction-stable doubly bottom tetraquarks on the lattice.
\emph{Phys. Rev. D} \textbf{2024}, \emph{110}, 094503. [\href{http://dx.doi.org/10.1103/PhysRevD.110.094503}{CrossRef}]

\bibitem{Sil93} Silvestre-Brac, B.; Semay, C.
Systematics of $L=0$ $q^2\bar q^2$ systems.
\emph{Z. Phys. C} \textbf{1993}, \emph{57}, 273--282. [\href{http://dx.doi.org/10.1007/BF01565058}{CrossRef}]

\bibitem{Ric18} Richard, J.-M.; Valcarce, A.; Vijande, J.
Few-body quark dynamics for doubly heavy baryons and tetraquarks.
\emph{Phys. Rev. C} \textbf{2018}, \emph{97}, 035211. [\href{http://dx.doi.org/10.1103/PhysRevC.97.035211}{CrossRef}]

\bibitem{Gar22} Garcilazo, H.; Valcarce, A.
Constituent quark-model hidden-flavor pentaquarks.
\emph{Phys. Rev. D} \textbf{2022}, \emph{105}, 114016. [\href{http://dx.doi.org/10.1103/PhysRevD.105.114016}{CrossRef}]

\bibitem{Gac22} Garcilazo, H.; Valcarce, A.
Hidden-flavor pentaquarks.
\emph{Phys. Rev. D} \textbf{2022}, \emph{106}, 114012. [\href{http://dx.doi.org/10.1103/PhysRevD.106.114012}{CrossRef}]

\bibitem{Gon18} Gongyo, S.; Sasaki, K.; Aoki, S.; Doi, T.;
Hatsuda, T.; Ikeda, Y.; Inoue, T.; Iritani, T.; Ishii, N.;
Miyamoto, T.; et al.
Most Strange Dibaryon from Lattice QCD.
\emph{Phys. Rev. Lett.} \textbf{2018}, \emph{120}, 212001. [\href{http://dx.doi.org/10.1103/PhysRevLett.120.212001}{CrossRef}]

\bibitem{Jua19} \textls[-15]{Junnarkar, P.; Mathur, N.
Deuteronlike Heavy Dibaryons from Lattice Quantum Chromodynamics.
\emph{Phys. Rev. Lett.} \textbf{2019}, \emph{123}, 162003. [\href{http://dx.doi.org/10.1103/PhysRevLett.123.162003}{CrossRef}]}

\bibitem{Lyu21} Lyu, Y.; Tong, H.; Sugiura, T.; Aoki, S.; Doi, T.;
Hatsuda, T.; Meng, J.; Miyamoto, T.
Dibaryon with Highest Charm Number near Unitarity from Lattice QCD.
\emph{Phys. Rev. Lett.} \textbf{2021}, \emph{127}, 072003. [\href{http://dx.doi.org/10.1103/PhysRevLett.127.072003}{CrossRef}]

\bibitem{Mat23} Mathur, N.; Padmanath, M.; Chakraborty, D.
Strongly Bound Dibaryon with Maximal Beauty Flavor from Lattice QCD.
\emph{Phys. Rev. Lett.} \textbf{2023}, \emph{130}, 111901. [\href{http://dx.doi.org/10.1103/PhysRevLett.130.111901}{CrossRef}]

\bibitem{Gac18} Garcilazo, H.; Valcarce, A.
$T_{bbb}$: A three $B$--meson bound state.
\emph{Phys. Lett. B} \textbf{2018}, \emph{784}, 169. [\href{http://dx.doi.org/10.1016/j.physletb.2018.07.055}{CrossRef}]

\bibitem{Mam19} Ma, L.; Wang, Q.; Meissner, U.-G.
Trimeson bound state $BBB^*$ via a delocalized $\pi$ bond.
\emph{Phys. Rev. D} \textbf{2019}, \emph{100}, 014028. [\href{http://dx.doi.org/10.1103/PhysRevD.100.014028}{CrossRef}]

\bibitem{Wul23} Wu, T.-W.; Luo, S.-Q.; Liu, M.-Z.; Geng, L.-S.; Liu, X.
Tribaryons with lattice QCD and one-boson exchange potentials.
\emph{Phys. Rev. D} \textbf{2023}, \emph{108}, L091506. [\href{http://dx.doi.org/10.1103/PhysRevD.108.L091506}{CrossRef}]

\bibitem{Gar24} Garcilazo, H.; Valcarce, A.
$\Omega_{bbb}\Omega_{bbb}\Omega_{bbb}$ tribaryons.
\emph{Rev. Mex. Fis.} \textbf{2024}, \emph{70}, 041202. [\href{http://dx.doi.org/10.31349/RevMexFis.70.041202}{CrossRef}]

\bibitem{Gar25} Garcilazo, H.; Valcarce, A.
Pauli principle forbids $\Omega_{bbb}\Omega_{bbb}\Omega_{bbb}$
bound states.
\emph{Phys. Rev. D} \textbf{2025}, \emph{111}, 014035. [\href{http://dx.doi.org/10.1103/PhysRevD.111.014035}{CrossRef}]

\bibitem{Ric20} Richard, J.-M.; Valcarce, A.; Vijande, J.
Very Heavy Flavored Dibaryons.
\emph{Phys. Rev. Lett.} \textbf{2020}, \emph{124}, 212001. [\href{http://dx.doi.org/10.1103/PhysRevLett.124.212001}{CrossRef}] [\href{http://www.ncbi.nlm.nih.gov/pubmed/32530666}{PubMed}]

\bibitem{Gar18} Garcilazo, H.; Valcarce, A.
Width of a two-body coupled-channel resonance.
\emph{Eur. Phys. J. C} \textbf{2018}, \emph{78}, 259. [\href{http://dx.doi.org/10.1140/epjc/s10052-018-5747-7}{CrossRef}]

\bibitem{Gai22} Garcilazo, H.; Valcarce, A.
$(I, J^P) = (1, 1/2^+)$ $\Sigma NN$ Quasibound State.
\emph{Symmetry} \textbf{2022}, \emph{14}, 2381. [\href{http://dx.doi.org/10.3390/sym14112381}{CrossRef}]

\bibitem{Aai15} Aaij, R.; Adeva, B.; Adinolfi, M.; Affolder, A.;
Ajaltouni, Z.; Akar, S.; Albrecht, J.; Alessio, F.;
Alexander, M.; Ali, S.; et al.
Observation of $J/\Psi p$ Resonances Consistent with Pentaquark States in $\Lambda^0 \to J/\Psi K^- p$ Decays.
\emph{Phys. Rev. Lett.} \textbf{2015}, \emph{115}, 072001. [\href{http://dx.doi.org/10.1103/PhysRevLett.115.072001}{CrossRef}]

\bibitem{Aai19}  {Aaij, R.; Abell\'an Beteta, C.; Adeva, B.;
Adinolfi, M.; Aidala, C.A.; Ajaltouni, Z.; Akar, S.; Albicocco, P.;
Albrecht, J.; Alessio, F.; et al.
Observation of a Narrow Pentaquark State $P_c(4312)^+$, and of the
Two-Peak Structure of the $P_c(4450)^+$.
\emph{Phys. Rev. Lett.} \textbf{2019}, \emph{122},~222001. [\href{http://dx.doi.org/10.1103/PhysRevLett.122.222001}{CrossRef}]}

\bibitem{Che22} Aaij, R.; Abdelmotteleb, A.S.W.; Abellan Beteta, C.;
Abudin\'en, F.; Ackernley, T.; Adeva, B.;
Adinolfi, M.; Adlarson, P.; Afsharnia, H.; Agapopoulou, C.; et al.
Observation of a $J/\Psi \Lambda$ Resonance Consistent with a Strange Pentaquark Candidate in $B^- \to J/\Psi \Lambda \bar p$ Decays.
\emph{Phys. Rev. Lett.} \textbf{2023}, \emph{131}, 031901. [\href{http://dx.doi.org/10.1103/PhysRevLett.131.031901}{CrossRef}] [\href{http://www.ncbi.nlm.nih.gov/pubmed/37540878}{PubMed}]

\bibitem{Aak21} Aaij, R.; Abell\'an Beteta, C.; Ackernley, T.;
Adeva, B.; Adinolfi, M.; Afsharnia, H.; Aidala, C.A.;
Aiola, S.; Ajaltouni, Z.; Akar, S.; \mbox{et al.}
Evidence of a $J/\Psi \Lambda$ structure and observation of excited
$\Xi^-$ states in the $\Xi_b^- \to J/\Psi \Lambda K^-$ decay.
\emph{Sci. Bull.} \textbf{2021}, \emph{66}, 1278--1287.

\bibitem{Don24} Dong, X.; Zou, S.M.; Zhang, H.Y.; Wang, X.L.;
Adachi, I.; Ahn, J.K.; Aihara, H.; Al Said, S.; Asner, D.M.;
Atmacan, H.; Ayad, R.; \mbox{et al.}
Search for a pentaquark state decaying into $p J/\Psi$ in $\Upsilon(1,2S)$ inclusive decays at Belle.
\emph{arXiv} \textbf{2024}, arXiv:2403.04340.

\bibitem{Ada25} Adachi, I.; Aggarwal, L.; Ahmed, H.; Ahn, J.K.;
Aihara, H.; Akopov, N.; Alhakami, M.; Aloisio, A.; Althubiti, N.;
Asner, D.M.; et al.
Search for $P_{c\bar c s}(4459)^0$ and $P_{c\bar c s}(4338)^0$ in $\Upsilon(1S,2S)$ inclusive decays at Belle.
\emph{arXiv} \textbf{2025}, arXiv:2502.09951. [\href{http://dx.doi.org/10.1103/pf8m-6j69}{CrossRef}]

\bibitem{Ans93} Anselmino, M.; Predazzi, E.; Ekelin, S.;
Fredriksson, S.; Lichtenberg, D.B.
Diquarks.
\emph{Rev. Mod. Phys.} \textbf{1993}, \emph{65}, 1199--1234. [\href{http://dx.doi.org/10.1103/RevModPhys.65.1199}{CrossRef}]

\bibitem{Fre82} Fredriksson, S.;  Jandel, M.
Diquark Deuteron.
\emph{Phys. Rev. Lett.} \textbf{1982}, \emph{48}, 14. [\href{http://dx.doi.org/10.1103/PhysRevLett.48.14}{CrossRef}]

\bibitem{Mai15} Maiani, L.; Polosa, A.D.; Riquer, V.
The New Pentaquarks in the Diquark Model.
\emph{Phys. Lett. B} \textbf{2015}, \emph{749}, 289--291. [\href{http://dx.doi.org/10.1016/j.physletb.2015.08.008}{CrossRef}]

\bibitem{Gir19} Giron, J.F.; Lebed, R.F.; Peterson, C.T.
The Dynamical Diquark Model: First Numerical Results.
\emph{J. High Energy Phys.} \textbf{2019}, \emph{05}, 061. [\href{http://dx.doi.org/10.1007/JHEP05(2019)061}{CrossRef}]

\bibitem{Ali19} Ali, A.; Ahmed, I.; Aslam, M.J.; Parkhomenko, A.Y.; Rehman, A.
Mass spectrum of the hidden-charm pentaquarks in the compact diquark model.
\emph{J. High Energy Phys.} \textbf{2019}, \emph{10}, 256. [\href{http://dx.doi.org/10.1007/JHEP10(2019)256}{CrossRef}]

\bibitem{Shi21} Shi, P.-P.; Huang, F.; Wang, W.-L.
Hidden charm pentaquark states in a diquark model.
\emph{Eur. Phys. J. A} \textbf{2021}, \emph{57}, 237. [\href{http://dx.doi.org/10.1140/epja/s10050-021-00542-4}{CrossRef}]

\bibitem{Wul17} Wu, J.; Liu, Y.-R.; Chen, K.; Liu, X.; Zhu, S.-L.
Hidden-charm pentaquarks and their hidden-bottom and $B_c$-like partner states.
\emph{Phys. Rev. D} \textbf{2017}, \emph{95}, 034002. [\href{http://dx.doi.org/10.1103/PhysRevD.95.034002}{CrossRef}]

\bibitem{Her20} Hern\'andez, E.; Vijande, J.; Valcarce, A.; Richard, J.-M.
Spectroscopy, lifetime and decay modes of the $T_{bb}^-$ tetraquark.
\emph{Phys. Lett. B} \textbf{2020}, \emph{800}, 135073. [\href{http://dx.doi.org/10.1016/j.physletb.2019.135073}{CrossRef}]

\bibitem{Men21} Meng, Q.; Hiyama, E.; Hosaka, A.; Oka, M.; Gubler, P.;
Can, K.U.; Takahashi, T.T.; Zong, H.S.
Stable double-heavy tetraquarks: spectrum and structure.
\emph{Phys. Lett. B} \textbf{2021}, \emph{814}, 136095. [\href{http://dx.doi.org/10.1016/j.physletb.2021.136095}{CrossRef}]

\bibitem{Sem94} Semay, C.; Silvestre-Brac, B.
Diquonia and potential models.
\emph{Z. Phys. C} \textbf{1994}, \emph{61}, 271--275. [\href{http://dx.doi.org/10.1007/BF01413104}{CrossRef}]

\bibitem{Jan04} Janc, D.; Rosina, M.
The $T_{cc} = DD^*$ Molecular State.
\emph{Few-Body Syst.} \textbf{2004}, \emph{35}, 175--196. [\href{http://dx.doi.org/10.1007/s00601-004-0068-9}{CrossRef}]

\bibitem{Ric17} Richard, J.-M.; Valcarce, A.; Vijande, J.
Stable heavy pentaquarks in constituent models.
\emph{Phys. Lett. B} \textbf{2017}, \emph{774}, 710--714. [\href{http://dx.doi.org/10.1016/j.physletb.2017.10.036}{CrossRef}]

\bibitem{Hiy18} Hiyama, E.; Hosaka, A.; Oka, M.; Richard, J.-M.
Quark model estimate of hidden-charm pentaquark resonances.
\emph{Phys. Rev. C} \textbf{2018}, \emph{98}, 045208. [\href{http://dx.doi.org/10.1103/PhysRevC.98.045208}{CrossRef}]

\bibitem{Meg19} Meng, Q.; Hiyama, E.; Can, K.U.; Gubler, P.; Oka, M.;
Hosaka, A.; Zong, H.
Compact $sss\bar c$ pentaquark states predicted by a quark model.
\emph{Phys. Lett. B} \textbf{2019}, \emph{798}, 135028. [\href{http://dx.doi.org/10.1016/j.physletb.2019.135028}{CrossRef}]

\bibitem{Sil96} Silvestre-Brac, B.
Spectrum and static properties of heavy baryons.
\emph{Few Body Syst.} \textbf{1996}, \emph{20}, 1--25. [\href{http://dx.doi.org/10.1007/s006010050028}{CrossRef}]

\bibitem{Ale11} Alex, A.; Kalus, M.; Huckleberry, A.; von Delft, J.
A Numerical algorithm for the explicit calculation of $SU(N)$ and $SL(N,C)$ Clebsch-Gordan coefficients.
\emph{J. Math. Phys.} \textbf{2011}, \emph{52}, 023507. [\href{http://dx.doi.org/10.1063/1.3521562}{CrossRef}]

\bibitem{Vij04} \textls[-15]{Vijande, J.; Fern\'andez, F.; Valcarce, A.; Silvestre-Brac, B.
Tetraquarks in a chiral constituent quark model.
\emph{Eur. Phys. J. A} \textbf{2004}, \emph{19}, 383. [\href{http://dx.doi.org/10.1140/epja/i2003-10128-9}{CrossRef}]}

\bibitem{Hua16} Huang, H.; Deng, C.; Ping, J.; Wang, F.
Possible pentaquarks with heavy quarks.
\emph{Eur. Phys. J. C} \textbf{2016}, \emph{76}, 624. [\href{http://dx.doi.org/10.1140/epjc/s10052-016-4476-z}{CrossRef}]

\bibitem{Yan17} Yang, G.; Ping, J.; Wang, F.
Structure of pentaquarks $P_c^+$ in the chiral quark model.
\emph{Phys. Rev. D} \textbf{2017}, \emph{95}, 014010. [\href{http://dx.doi.org/10.1103/PhysRevD.95.014010}{CrossRef}]

\bibitem{Oka80} Oka, M.; Yazaki, K.
Nuclear Force in a Quark Model.
\emph{Phys. Lett. B} \textbf{1980}, \emph{90}, 41--44. [\href{http://dx.doi.org/10.1016/0370-2693(80)90046-5}{CrossRef}]

\bibitem{Gol89} \textls[-15]{Goldman, T.; Maltman, K.; Stephenson, G.J.;
Schmidt, K.E., Jr.; Wang, F.
``Inevitable'' nonstrange dibaryon.
\emph{Phys. Rev. C} \textbf{1989}, \emph{39}, 1889. [\href{http://dx.doi.org/10.1103/PhysRevC.39.1889}{CrossRef}]}

\bibitem{Val01} Valcarce, A.; Garcilazo, H.; Mota, R.D.; Fern\'andez, F.
$\Delta\Delta$ and $\Delta\Delta\Delta$ bound states.
\emph{J. Phys. G} \textbf{2001}, \emph{27}, L1--L7. [\href{http://dx.doi.org/10.1088/0954-3899/27/1/101}{CrossRef}]

\bibitem{Val05} Valcarce, A.; Garcilazo, H.; Fern\'andez, F.; Gonz\'alez, P.
Quark-model study of few-baryon systems.
\emph{Rep. Prog. Phys.} \textbf{2005}, \emph{68}, 965--1042. [\href{http://dx.doi.org/10.1088/0034-4885/68/5/R01}{CrossRef}]

\bibitem{Pan01} Pang, H.R.; Ping, J.L.; Wang, F.; Goldman, T.
Phenomenological study of hadron interaction models.
\emph{Phys. Rev. C} \textbf{2001}, \emph{65}, 014003. [\href{http://dx.doi.org/10.1103/PhysRevC.65.014003}{CrossRef}]

\bibitem{Saz22} Sazdjian, H.
The Interplay between Compact and Molecular Structures in Tetraquarks.
\emph{Symmetry} \textbf{2022}, \emph{14}, 515. [\href{http://dx.doi.org/10.3390/sym14030515}{CrossRef}]

\bibitem{Kar17} Karliner, M.; Rosner, J.L.
Discovery of the Doubly Charmed $\Xi_{cc}$ Baryon Implies a Stable $bb\bar u \bar d$ Tetraquark.
\emph{Phys. Rev. Lett.} \textbf{2017}, \emph{119}, 202001. [\href{http://dx.doi.org/10.1103/PhysRevLett.119.202001}{CrossRef}]

\bibitem{Eic17} Eichten, E.J.; Quigg, C.
Heavy-Quark Symmetry Implies Stable Heavy Tetraquark Mesons $Q_iQ_j\bar q_k \bar q_l$.
\emph{Phys. Rev. Lett.} \textbf{2017}, \emph{119}, 202002. [\href{http://dx.doi.org/10.1103/PhysRevLett.119.202002}{CrossRef}] [\href{http://www.ncbi.nlm.nih.gov/pubmed/29219332}{PubMed}]

\bibitem{Fad61} Faddeev, L.D.
Scattering Theory for a Three-Particle System.
\emph{Sov. Phys. JETP} \textbf{1961}, \emph{12}, 1014--1019.

\bibitem{Fad65} Faddeev, L.D.
{\it Mathematical Aspects of the Three-Body Problem
in Quantum Scattering Theory};
Daley: New York, NY, USA, 1965.

\bibitem{Gar03} Garcilazo, H.
Momentum-space Faddeev calculations for confining potentials.
\emph{Phys. Rev. C} \textbf{2003}, \emph{67}, 055203. [\href{http://dx.doi.org/10.1103/PhysRevC.67.055203}{CrossRef}]

\bibitem{Fra21} Francis, A.; de Forcrand, P.; Lewis, R.; Maltman, K.
Diquark properties from full QCD lattice simulations.
\emph{J. High Energy Phys.} \textbf{2022}, \emph{05}, 062. [\href{http://dx.doi.org/10.1007/JHEP05(2022)062}{CrossRef}]

\bibitem{Ale06} Alexandrou, C.; de Forcrand, P.; Lucini, B.
Searching for diquarks in hadrons.
\emph{Proc. Sci.} \textbf{2006}, \emph{053}, LAT2005.

\bibitem{Gre10} Green, J.; Negele, J.; Engelhardt, M.; Varilly, P.
Spatial diquark correlations in a hadron.
\emph{Proc. Sci. Lattice} \textbf{2010}, \emph{2010}, 140.

\bibitem{Wag20} Wang, B.; Meng, L.; Zhu, S.-L.
Spectrum of the strange hidden charm molecular pentaquarks in chiral effective field theory.
\emph{Phys. Rev. D} \textbf{2020}, \emph{101}, 034018. [\href{http://dx.doi.org/10.1103/PhysRevD.101.034018}{CrossRef}]

\bibitem{Hup22} Hu, X.; Ping, J.
Investigation of hidden-charm pentaquarks with strangeness $S=-1$.
\emph{Eur. Phys. J. C} \textbf{2022}, \emph{82}, 118. [\href{http://dx.doi.org/10.1140/epjc/s10052-022-10047-z}{CrossRef}]

\bibitem{Fer20} Ferretti, J.; Santopinto, E.
Hidden-charm and bottom tetra- and pentaquarks with strangeness in the hadro-quarkonium and compact tetraquark models.
\emph{J. High Energy Phys.} \textbf{2020}, \emph{04}, 119. [\href{http://dx.doi.org/10.1007/JHEP04(2020)119}{CrossRef}]

\bibitem{Yan19} Yang, G.; Ping, J.; Segovia, J.
Hidden-bottom pentaquarks.
\emph{Phys. Rev. D} \textbf{2019}, \emph{99}, 014035. [\href{http://dx.doi.org/10.1103/PhysRevD.99.014035}{CrossRef}]

\bibitem{Fer19} Ferretti, J.; Santopinto, E.; Anwar, M.N.; Bedolla, M.A.
The baryo-quarkonium picture for hidden-charm and bottom pentaquarks and LHCb $P_c(4380)$ and $P_c(4450)$ states.
\emph{Phys. Lett. B} \textbf{2019}, \emph{789}, 562--567. [\href{http://dx.doi.org/10.1016/j.physletb.2018.09.047}{CrossRef}]

\bibitem{Zhu16} Zhu, R.; Qiao, C.-F.
Pentaquark states in a diquark–triquark model.
\emph{Phys. Lett. B} \textbf{2016}, \emph{756}, 259--264. [\href{http://dx.doi.org/10.1016/j.physletb.2016.03.022}{CrossRef}]

\bibitem{Wun10} Wu, J.-J.; Molina, R.; Oset, E.;  Zou, S.B.
Prediction of narrow $N^*$ and $\Lambda^*$ resonances with hidden charm above 4 GeV.
\emph{Phys. Rev. Lett.} \textbf{2010}, \emph{105}, 232001. [\href{http://dx.doi.org/10.1103/PhysRevLett.105.232001}{CrossRef}] [\href{http://www.ncbi.nlm.nih.gov/pubmed/21231450}{PubMed}]

\bibitem{Wan11} Wang, W.L.; Huang, F.; Zhang, Z.Y.; Zou, B.S.
$\Sigma_c \bar D$ and $\Lambda_c \bar D$ states in a chiral quark model.
\emph{Phys. Rev. C} \textbf{2011}, \emph{84}, 015203. [\href{http://dx.doi.org/10.1103/PhysRevC.84.015203}{CrossRef}]

\bibitem{Yan12} Yang, Z.-C.; Sun, Z.-F.; He, J.; Liu, X.; Zhu, S.-L.
The possible hidden-charm molecular baryons composed of anti-charmed meson and charmed baryon.
\emph{Chin. Phys. C} \textbf{2012}, \emph{36}, 6--13. [\href{http://dx.doi.org/10.1088/1674-1137/36/1/002}{CrossRef}]

\bibitem{Wul12} Wu, J.-J.; Lee, T.-S.H.; Zou, B.S.
Nucleon resonances with hidden charm in coupled-channels models.
\emph{Phys. Rev. C} \textbf{2012}, \emph{85}, 044002. [\href{http://dx.doi.org/10.1103/PhysRevC.85.044002}{CrossRef}]

\bibitem{Xia13} Xiao, C.W.; Nieves, J.; Oset, E.
Combining heavy quark spin and local hidden gauge symmetries in the dynamical generation of hidden charm baryons.
\emph{Phys. Rev. D} \textbf{2013}, \emph{88}, 056012. [\href{http://dx.doi.org/10.1103/PhysRevD.88.056012}{CrossRef}]

\bibitem{Yaa17} Yamaguchi, Y.; Giachino, A.; Hosaka, A.; Santopinto, E.;
Takeuchi, S.; Takizawa, M.
Hidden-charm and bottom meson-baryon molecules coupled with five-quark states.
\emph{Phys. Rev. D} \textbf{2017}, \emph{96}, 114031. [\href{http://dx.doi.org/10.1103/PhysRevD.96.114031}{CrossRef}]

\bibitem{Wum11} Wu, J.-J.; Molina, R.; Oset, E.; Zou, B.S.
Dynamically generated $N^*$ and $\Lambda^*$ resonances in the hidden charm sector around 4.3 GeV.
\emph{Phys. Rev. C} \textbf{2011}, \emph{84}, 015202. [\href{http://dx.doi.org/10.1103/PhysRevC.84.015202}{CrossRef}]

\bibitem{Eid16} \textls[-15]{Eides, M.I.; Petrov, V.Y.; Polyakov, M.V.
Narrow nucleon-$\Psi(2S)$ bound state and LHCb pentaquarks.
\emph{Phys. Rev. D} \textbf{2016}, \emph{93}, 054039. [\href{http://dx.doi.org/10.1103/PhysRevD.93.054039}{CrossRef}]}

\bibitem{Kar15} Karliner, M.; Rosner, J.L.
New Exotic Meson and Baryon Resonances from Doubly Heavy Hadronic Molecules.
\emph{Phys. Rev. Lett.} \textbf{2015}, \emph{115}, 122001. [\href{http://dx.doi.org/10.1103/PhysRevLett.115.122001}{CrossRef}]

\bibitem{Wag21} Wang, J.-Z.; Liu, X.; Matsuki, T.
Evidence supporting the existence of $P_c(4380)^\pm$ from the recent measurements of $B_s \to J/\Psi p \bar p$.
\emph{Phys. Rev. D} \textbf{2021}, \emph{104}, 114020. [\href{http://dx.doi.org/10.1103/PhysRevD.104.114020}{CrossRef}]

\bibitem{Bro90} Brodsky, S.J.; Schmidt, I.; de Teramond, G.F.
Nuclear-bound quarkonium.
\emph{Phys. Rev. Lett.} \textbf{1990}, \emph{64}, 1011. [\href{http://dx.doi.org/10.1103/PhysRevLett.64.1011}{CrossRef}]

\bibitem{Xia19} Xiao, C.-J.; Huang, Y.; Dong, Y.-B.; Geng, L.-S.; Chen, D.-Y.
Exploring the molecular scenario of $P_c(4312)$, $P_c(4440)$ and $P_c(4457)$.
\emph{Phys. Rev. D} \textbf{2019}, \emph{100}, 014022. [\href{http://dx.doi.org/10.1103/PhysRevD.100.014022}{CrossRef}]

\bibitem{She19} \textls[-15]{Shen, C.-W.; Wu, J.-J.; Zou, B.-S.
Decay behaviors of possible $\Lambda_{c \bar c}$ states in hadronic molecule pictures.
\emph{Phys. Rev. D} \textbf{2019}, \emph{100}, 056006. [\href{http://dx.doi.org/10.1103/PhysRevD.100.056006}{CrossRef}]}

\bibitem{Wan20} Wang, F.-L.; Chen, R.; Liu, Z.-W.; Liu, X.
Probing new types of $P_c$ states inspired by the interaction between an $S$-wave charmed baryon and an anticharmed meson in a $\bar T$ doublet state.
\emph{Phys. Rev. C} \textbf{2020}, \emph{101}, 025201. [\href{http://dx.doi.org/10.1103/PhysRevC.101.025201}{CrossRef}]

\bibitem{Bur15} Burns, T.J.
Phenomenology of $P_c(4380)^+$, $P_c(4450)^+$ and related states.
\emph{Eur. Phys. J. A} \textbf{2015}, \emph{51}, 152. [\href{http://dx.doi.org/10.1140/epja/i2015-15152-6}{CrossRef}]

\bibitem{Per16} Perevalova, I.A.; Polyakov, M.V.; Schweitzer, P.
LHCb pentaquarks as a baryon-$\Psi(2S)$ bound state: Prediction of isospin-$3/2$ pentaquarks with hidden charm.
\emph{Phys. Rev. D} \textbf{2016}, \emph{94}, 054024. [\href{http://dx.doi.org/10.1103/PhysRevD.94.054024}{CrossRef}]

\bibitem{Wen19} Weng, X.-Z.; Chen, X.-L.; Deng, W.-Z.; Zhu, S.-L.
Hidden-charm pentaquarks and $P_c$ states.
\emph{Phys. Rev. D} \textbf{2019}, \emph{100}, 016014. [\href{http://dx.doi.org/10.1103/PhysRevD.100.016014}{CrossRef}]

\bibitem{Mai14} Maiani, L.; Piccinini, F.; Polosa, A.D.; Riquer, V.
$Z(4430)$ and a new paradigm for spin interactions in tetraquarks.
\emph{Phys. Rev. D} \textbf{2014}, \emph{89}, 114010. [\href{http://dx.doi.org/10.1103/PhysRevD.89.114010}{CrossRef}]

\bibitem{Azi17} Azizi, K.; Sarac, Y.; Sundu, H.
Analysis of $P_c^+(4380)$ andd $P_c^+(4450)$ as pentaquark states in the molecular picture with QCD sum rules.
\emph{Phys. Rev. D} \textbf{2017}, \emph{95}, 094016. [\href{http://dx.doi.org/10.1103/PhysRevD.95.094016}{CrossRef}]

\bibitem{Che19} Chen, R.; Sun, Z.-F.; Liu, X.; Zhu, S.-L.
Strong LHCb evidence supporting the existence of the hidden-charm molecular pentaquarks.
\emph{Phys. Rev. D} \textbf{2019}, \emph{100}, 011502. [\href{http://dx.doi.org/10.1103/PhysRevD.100.011502}{CrossRef}]

\bibitem{Zha19} Zhang, J.-R.
Exploring $\Sigma_c \bar D$ state: With focus on $P_c(4312)^+$.
\emph{Eur. Phys. J. C} \textbf{2019}, \emph{79}, 1001. [\href{http://dx.doi.org/10.1140/epjc/s10052-019-7529-2}{CrossRef}]

\bibitem{Wan21} Wang, Z.-G.; Xin, Q.
Analysis of hidden-charm pentaquark molecular states with and without strangeness via the QCD sum rules.
\emph{Chin. Phys. C} \textbf{2021}, \emph{45}, 123105. [\href{http://dx.doi.org/10.1088/1674-1137/ac2a1d}{CrossRef}]

\bibitem{Pim20} Pimikov, A.; Lee, H.-J.; Zhang, P.
Hidden-charm pentaquarks with color-octet substructure in QCD sum rules.
\emph{Phys. Rev. D} \textbf{2020}, \emph{101}, 014002. [\href{http://dx.doi.org/10.1103/PhysRevD.101.014002}{CrossRef}]

\bibitem{Nar21} Narison, S.
Modern status of heavy quark sum rules in QCD.
\emph{Nucl. Part. Phys. Proc.} \textbf{2021}, \emph{312--317}, 87--93. [\href{http://dx.doi.org/10.1016/j.nuclphysbps.2021.05.023}{CrossRef}]

\bibitem{Wan19} Wang, B.; Meng, L.; Zhu, S.-L.
Hidden-charm and hidden-bottom molecular pentaquarks in chiral effective field theory.
\emph{J. High Energy Phys.} \textbf{2019}, \emph{11}, 108. [\href{http://dx.doi.org/10.1007/JHEP11(2019)108}{CrossRef}]

\bibitem{Men19} Meng, L.; Wang, B.; Wang, G.-J.; Zhu, S.-L.
Hidden charm pentaquark states and $\Sigma_c \bar D^{(*)}$ interaction in chiral perturbation theory.
\emph{Phys. Rev. D} \textbf{2019}, \emph{100}, 014031. [\href{http://dx.doi.org/10.1103/PhysRevD.100.014031}{CrossRef}]

\bibitem{Yam17} Yamaguchi, Y.; Santopinto, E.
Hidden-charm pentaquarks as a meson-baryon molecule with coupled channels for $\bar D^{(*)} \Lambda_c$ and $\bar D^{(*)} \Sigma_c^{(*)}$.
\emph{Phys. Rev. D} \textbf{2017}, \emph{96}, 014018. [\href{http://dx.doi.org/10.1103/PhysRevD.96.014018}{CrossRef}]

\bibitem{Yal21} Yalikun, N.; Lin, Y.-H.; Guo, F.-K.; Kamiya, Y.; Zou, B.-S.
Coupled-channel effects of the $\Sigma_c^{(*)} \bar D^{(*)} - \Lambda_c(2595) \bar D$ system and molecular nature of the $P_c$ pentaquark states from one-boson exchange model.
\emph{Phys. Rev. D} \textbf{2021}, \emph{104}, 094039. [\href{http://dx.doi.org/10.1103/PhysRevD.104.094039}{CrossRef}]

\bibitem{Yan24} Yang, G.; Ping, J.; Segovia, J.
Hidden-Charm Pentaquarks with Strangeness in a Chiral Quark Model.
\emph{Symmetry} \textbf{2024}, \emph{16}, 354. [\href{http://dx.doi.org/10.3390/sym16030354}{CrossRef}]

\bibitem{Ric19} Richard, J.-M.; Valcarce, A.; Vijande, J.
Pentaquarks with anticharm or beauty revisited.
\emph{Phys. Lett. B} \textbf{2019}, \emph{790}, 248--250. [\href{http://dx.doi.org/10.1016/j.physletb.2019.01.031}{CrossRef}]

\bibitem{Dub08} Dubynskiy, S.; Voloshin, M.B.
Hadro-Charmonium.
\emph{Phys. Lett. B} \textbf{2008}, \emph{666}, 344--346. [\href{http://dx.doi.org/10.1016/j.physletb.2008.07.086}{CrossRef}]

\bibitem{Tsu11} Tsushima, K.; Lu, D.H.; Krein, G.; Thomas, A.W.
$J/\Psi$-nuclear bound states.
\emph{Phys. Rev. C} \textbf{2011}, \emph{83}, 065208. [\href{http://dx.doi.org/10.1103/PhysRevC.83.065208}{CrossRef}]

\bibitem{Cob20} Cobos-Mart\'\i nez, J.J.; Tsushima, K.; Krein, G.; Thomas, A.W.
$\eta_c$-nucleus bound states.
\emph{Phys. Lett. B} \textbf{2020}, \emph{811}, 135882. [\href{http://dx.doi.org/10.1016/j.physletb.2020.135882}{CrossRef}]

\bibitem{Ric93} Richard, J.-M.; Fr\"ohlich, J.; Graf, G.-M.; Seifert, M.
Proof of stability of the hydrogen molecule.
\emph{Phys. Rev. Lett.} \textbf{1993}, \emph{71}, 1332. [\href{http://dx.doi.org/10.1103/PhysRevLett.71.1332}{CrossRef}] [\href{http://www.ncbi.nlm.nih.gov/pubmed/10055513}{PubMed}]

\bibitem{Fra17} Francis, A.; Hudspith, R.J.; Lewis, R.; Maltman, K.
Lattice Prediction for Deeply Bound Doubly Heavy Tetraquarks.
\emph{Phys. Rev. Lett.} \textbf{2017}, \emph{118}, 142001. [\href{http://dx.doi.org/10.1103/PhysRevLett.118.142001}{CrossRef}] [\href{http://www.ncbi.nlm.nih.gov/pubmed/28430484}{PubMed}]

\bibitem{Bic16} \textls[-15]{Bicudo, P.; Cichy, K.; Peters, A.; Wagner, M.
$BB$ interactions with static bottom quarks from lattice QCD.
\emph{Phys. Rev. D} \textbf{2016}, \emph{93}, 034501. [\href{http://dx.doi.org/10.1103/PhysRevD.93.034501}{CrossRef}]}

\bibitem{Jun19} Junnarkar, P.; Mathur, N.; Padmanath, M.
Study of doubly heavy tetraquarks in lattice QCD.
\emph{Phys. Rev. D} \textbf{2019}, \emph{99}, 034507. [\href{http://dx.doi.org/10.1103/PhysRevD.99.034507}{CrossRef}]

\bibitem{Luo17} \textls[-15]{Luo, S.-Q.; Chen, K.; Liu, X.; Liu, Y.-R.; Zhu, S.-L.
Exotic tetraquark states with the $qq\bar Q \bar Q$ configuration.
\emph{Eur. Phys. J. C} \textbf{2017}, \emph{77}, 709. [\href{http://dx.doi.org/10.1140/epjc/s10052-017-5297-4}{CrossRef}]}

\bibitem{Duc13} Du, M.-L.; Chen, W.; Chen, X.-L.; Zhu, S.-L.
Exotic $QQ\bar q \bar q$, $QQ\bar q \bar s$, $QQ\bar s \bar s$ states.
\emph{Phys. Rev. D} \textbf{2013}, \emph{87}, 014003. [\href{http://dx.doi.org/10.1103/PhysRevD.87.014003}{CrossRef}]

\bibitem{Cza18} Czarnecki, A.; Leng, B.; Voloshin, M.B.
Stability of tetrons.
\emph{Phys. Lett. B} \textbf{2018}, \emph{778}, 233--238. [\href{http://dx.doi.org/10.1016/j.physletb.2018.01.034}{CrossRef}]

\bibitem{Vij09} Vijande, J.; Valcarce, A.; Barnea, N.
Exotic meson-meson molecules and compact four-quark states.
\emph{Phys. Rev. D} \textbf{2009}, \emph{79}, 074010. [\href{http://dx.doi.org/10.1103/PhysRevD.79.074010}{CrossRef}]

\bibitem{Har81} \textls[-15]{Harvey, M.
On the Fractional Parentage Expansions of Color Singlet Six Quark States in a Cluster Model.
\emph{Nucl. Phys.} \textbf{1981}, \emph{352}, 301. [\href{http://dx.doi.org/10.1016/0375-9474(81)90412-7}{CrossRef}]}

\bibitem{Via09} Vijande, J.; Valcarce, A.
Probabilities in nonorthogonal bases: Four-quark systems.
\emph{Phys. Rev. C} \textbf{2009}, \emph{80}, 035204. [\href{http://dx.doi.org/10.1103/PhysRevC.80.035204}{CrossRef}]

\bibitem{Car12} Caram\'es, T.F.; Valcarce, A.; Vijande, J.
Too many $X's$, $Y's$ and $Z's$?
\emph{Phys. Lett. B} \textbf{2012}, \emph{709}, 358--361. [\href{http://dx.doi.org/10.1016/j.physletb.2012.02.020}{CrossRef}]

\bibitem{Ike14} Ikeda, Y.; Charron, B.; Aoki, S.; Doi, T.; Hatsuda, T.;
Inoue, T.; Ishii, N.; Murano, K.; Nemura, H.; Sasaki, K.
Charmed tetraquarks $T_{cc}$ and $T_{cs}$ from dynamical lattice QCD simulations.
\emph{Phys. Lett. B} \textbf{2014}, \emph{729}, 85--90. [\href{http://dx.doi.org/10.1016/j.physletb.2014.01.002}{CrossRef}]

\bibitem{Tor91} T\"ornqvist, N.A.
Possible large deuteronlike meson-meson states bound by pions.
\emph{Phys. Rev. Lett.} \textbf{1991}, \emph{67}, 556. [\href{http://dx.doi.org/10.1103/PhysRevLett.67.556}{CrossRef}]

\bibitem{Man93} Manohar, A.V.; Wise, M.B.
Exotic $QQ\bar q\bar q$ states in QCD.
\emph{Nucl. Phys. B} \textbf{1993}, \emph{399}, 17--33. [\href{http://dx.doi.org/10.1016/0550-3213(93)90614-U}{CrossRef}]

\bibitem{Eri93} Ericson, T.E.O.; Karl, G.
Strength of pion exchange in hadronic molecules.
\emph{Phys. Lett. B} \textbf{1993}, \emph{309}, 426--430. [\href{http://dx.doi.org/10.1016/0370-2693(93)90957-J}{CrossRef}]

\bibitem{Clo10} Close, F.; Downum, C.; Thomas, C.E.
Novel charmonium and bottomonium spectroscopies due to deeply bound hadronic molecules from single pion exchange.
\emph{Phys. Rev. D} \textbf{2010}, \emph{81}, 074033. [\href{http://dx.doi.org/10.1103/PhysRevD.81.074033}{CrossRef}]

\bibitem{Jur73} M. Juri\v{c}; Bohm, G.; Klabuhn, J.; Krecker, U.;
Wysotzki, F.; Coremans-Bertrand, G.; Sacton, J.; Wilquet, G.;
Cantwell, T.; Esmael, F.; et al.
A new determination of the binding-energy values of the light hypernuclei $(A \leq 15)$.
\emph{Nucl. Phys. B} \textbf{1973}, \emph{52}, 1--30. [\href{http://dx.doi.org/10.1016/0550-3213(73)90084-9}{CrossRef}]

\bibitem{Ess15} Esser, A.; Nagao, S.; Schulz, F.; Achenbach, P.;
Ayerbe Gayoso, C.; B\"ohm, R.; Borodina, O.; Bosnar, D.; Bozkurt, V.;
Debenjak, L.; \mbox{et al.}
Observation of $^4_\Lambda$H Hyperhydrogen by Decay-Pion Spectroscopy in Electron Scattering.
\emph{Phys. Rev. Lett.} \textbf{2015}, \emph{114}, 232501. [\href{http://dx.doi.org/10.1103/PhysRevLett.114.232501}{CrossRef}]

\bibitem{Gar17} Garcilazo, H.; Valcarce, A.; Caram\'es, T.F.
Three-body systems with open flavor heavy mesons.
\emph{Phys. Rev. D} \textbf{2017}, \emph{96}, 074009. [\href{http://dx.doi.org/10.1103/PhysRevD.96.074009}{CrossRef}]

\bibitem{Mar08} Mart\'\i nez Torres, A.; Khemchandani, K.P.; Oset, E.
Three-body resonances in two-meson–one-baryon systems.
\emph{Phys. Rev. C} \textbf{2008}, \emph{77}, 042203. [\href{http://dx.doi.org/10.1103/PhysRevC.77.042203}{CrossRef}]

\bibitem{Mar20} Mart\'\i nez Torres, A.; Khemchandani, K.P.; Roca, L.; Oset, E.
Few-body systems consisting of mesons.
\emph{Few Body Syst.} \textbf{2020}, \emph{61}, 35. [\href{http://dx.doi.org/10.1007/s00601-020-01568-y}{CrossRef}]

\bibitem{Zha17} Shah, Z.; Kumar-Rai, A.
Masses and Regge trajectories of triply heavy $\Omega_{ccc}$ and $\Omega_{bbb}$ baryons.
\emph{Eur. Phys. J. A} \textbf{2017}, \emph{53}, 195. [\href{http://dx.doi.org/10.1140/epja/i2017-12386-2}{CrossRef}]

\bibitem{Lea89} Leandri, J.; Silvestre-Brac, B.
Systematics of $Q\bar q^4$ systems with a pure chromomagnetic interaction.
\emph{Phys. Rev. D} \textbf{1989}, \emph{40}, 2340. [\href{http://dx.doi.org/10.1103/PhysRevD.40.2340}{CrossRef}]

\bibitem{Wul19} Wu, T.-W.; Liu, M.-Z.; Geng, L.-S.
Hiyama, E.; Pavon Valderrama, M.
$DK$, $DDK$, and $DDDK$ molecules–understanding the nature of the $D_{s0}^*(2317)$.
\emph{Phys. Rev. D} \textbf{2019}, \emph{100}, 034029.

\bibitem{Ort24} Ortega, P.G.
Exploring the Efimov effect in the $D^*D^*D^*$ system.
\emph{Phys. Rev. D} \textbf{2024}, \emph{110}, 034015. [\href{http://dx.doi.org/10.1103/PhysRevD.110.034015}{CrossRef}]

\bibitem{Gar97} Garcilazo, H.; Fern\'andez, F.; Valcarce, A.; Mota, R.D.
Bound states of $\Delta\Delta$ and $\Delta\Delta\Delta$ systems.
\emph{Phys. Rev. C} \textbf{1997}, \emph{56}, 84. [\href{http://dx.doi.org/10.1103/PhysRevC.56.84}{CrossRef}]

\bibitem{Gar87} Garcilazo, H.
Nonexistence of $\Lambda NN$ and $\Sigma NN$ bound states.
\emph{J. Phys. G} \textbf{1987}, \emph{13}, L63--L67. [\href{http://dx.doi.org/10.1088/0305-4616/13/5/002}{CrossRef}]

\bibitem{Oka87} Oka, M.; Shimizu, K.; Yazaki, K.
Hyperon-Nucleon and Hyperon-Hyperon Interaction in a Quark Model.
\emph{Nucl. Phys. A} \textbf{1987}, \emph{464}, 700--716. [\href{http://dx.doi.org/10.1016/0375-9474(87)90371-X}{CrossRef}]

\bibitem{Lib77} Liberman, D.A.
Short-range part of the nuclear force.
\emph{Phys. Rev. D} \textbf{1977}, \emph{16}, 1542. [\href{http://dx.doi.org/10.1103/PhysRevD.16.1542}{CrossRef}]

\bibitem{Oka84} Oka, M.; Yazaki, K.
Baryon baryon interaction from quark model viewpoint.
\emph{Int. Rev. Nucl. Phys.} \textbf{1984}, \emph{1}, 489--567.

\bibitem{Oka81} Oka, M.; Yazaki, K.
Short Range Part of Baryon Baryon Interaction in a Quark Model. 1. Formulation.
\emph{Prog. Theor. Phys.} \textbf{1981}, \emph{66}, 556--571. [\href{http://dx.doi.org/10.1143/PTP.66.556}{CrossRef}]

\bibitem{Oky81}  Oka, M.; Yazaki, K.
Short Range Part of Baryon Baryon Interaction in a Quark Model. 2. Numerical Results for $S$-Wave
\emph{Prog. Theor. Phys.} \textbf{1981}, \emph{66}, 572--587. [\href{http://dx.doi.org/10.1143/PTP.66.556}{CrossRef}]

\bibitem{Gar19} Garcilazo, H.; Valcarce, A.; Caram\'es, T.F.
Charmed baryon–nucleon interaction.
\emph{Eur. Phys. J. C} \textbf{2019}, \emph{79}, 598. [\href{http://dx.doi.org/10.1140/epjc/s10052-019-7110-z}{CrossRef}]

\bibitem{Gal14} Gal, A.; Garcilazo, H.; Valcarce, A.; Fern\'andez-Caram\'es, T.
Pion-assisted charmed dibaryon candidate.
\emph{Phys. Rev. D} \textbf{2014}, \emph{90},~014019. [\href{http://dx.doi.org/10.1103/PhysRevD.90.014019}{CrossRef}]

\bibitem{Buc12} Buchoff, M.I.; Luu, T.C.; Wasem, J.
$S$-wave scattering of strangeness $-3$ baryons.
\emph{Phys. Rev. D} \textbf{2012}, \emph{85}, 094511. [\href{http://dx.doi.org/10.1103/PhysRevD.85.094511}{CrossRef}]

\bibitem{Liu19} Liu, Y.-R.; Chen, H.-X.; Chen, W.; Liu, X.; Zhu, S.-L.
Pentaquark and Tetraquark states.
\emph{Prog. Part. Nucl. Phys.} \textbf{2019}, \emph{107}, 237--320. [\href{http://dx.doi.org/10.1016/j.ppnp.2019.04.003}{CrossRef}]

\bibitem{Wei90} Weinstein, J.D.; Isgur, N.
$K \bar K$ molecules.
\emph{Phys. Rev. D} \textbf{1990}, \emph{41}, 2236. [\href{http://dx.doi.org/10.1103/PhysRevD.41.2236}{CrossRef}]

\bibitem{Nav24} Navas, S.; Amsler, C.; Gutsche, T.; Hanhart, C.;
Hern\'andez-Rey, J.J.; Louren\c{c}o, C.; Masoni, A.; Mikhasenko, M.;
Mitchell, R.E.: Patrignani, C.; et al.  Review of Particle Physics.
\emph{Phys. Rev. D} \textbf{2024}, \emph{110}, 030001. [\href{http://dx.doi.org/10.1103/PhysRevD.110.030001}{CrossRef}]

\bibitem{Bre36} Breit, G.; Wigner, E.
Capture of Slow Neutrons.
\emph{Phys. Rev.} \textbf{1936}, \emph{49}, 519. [\href{http://dx.doi.org/10.1103/PhysRev.49.519}{CrossRef}]

\bibitem{Cec08} Ceci, S.; \v{S}varc, A.; Zauner, B.; Manley, D.M.; Capstick, S.
Model-independent resonance parameter extraction using the trace of K and T matrices.
\emph{Phys. Lett. B} \textbf{2008}, \emph{659}, 228--233. [\href{http://dx.doi.org/10.1016/j.physletb.2007.11.003}{CrossRef}]


\bibitem{Cec13}  {Ceci, S.; Korolija, M.; Zauner, B.} Model-Independent Extraction of the Pole and Breit-Wigner Resonance Parameters.
\emph{Phys. Rev. Lett.} \textbf{2013}, \emph{111}, 112004. [\href{http://dx.doi.org/10.1103/PhysRevLett.111.112004}{CrossRef}] [\href{http://www.ncbi.nlm.nih.gov/pubmed/24074077}{PubMed}]

\bibitem{Kar21} Karliner, M.; Rosner, J.L.
Strange pentaquarks and excited $\Xi$ hyperons in $\Xi_b^- \to J/\Psi \Lambda K^-$ final states.
\emph{Sci. Bull.} \textbf{2021}, \emph{66}, 1256. [\href{http://dx.doi.org/10.1016/j.scib.2021.04.013}{CrossRef}]
\end{thebibliography}
\end{document}